\documentclass[prx,twocolumn,amsmath,amssymb,showpacs,longbibliography]{revtex4-1}

\usepackage{hyperref}
\pdfoutput=1
\usepackage{graphicx}

\usepackage{amsmath}
\usepackage{bm}
\usepackage{amssymb}
\usepackage{textcomp}
\usepackage{stmaryrd} 
\usepackage{ulem}
\usepackage{gensymb}

\usepackage{import}
\usepackage{color}
\usepackage{verbatim}

\usepackage{float}

\usepackage{sistyle} 

\newcommand{\tr}[1]{{\textrm{#1}}}

\newcommand\abs[1]{\left| #1 \right|}

\def\XXint#1#2#3{{\setbox0=\hbox{$#1{#2#3}{\int}$}
     \vcenter{\hbox{$#2#3$}}\kern-.5\wd0}}

\begin{document}

\title{Charge versus energy transfer in atomically-thin\\ graphene-transition metal dichalcogenide van der Waals heterostructures}

\author{Guillaume Froehlicher}
\altaffiliation{Present address: Department of Physics, University of Basel, Klingelbergstrasse 82, CH-4056 Basel, Switzerland.}
\affiliation{Universit\'e de Strasbourg, CNRS, Institut de Physique et Chimie des Mat\'eriaux de Strasbourg (IPCMS), UMR 7504, F-67000 Strasbourg, France}

\author{Etienne Lorchat}
\affiliation{Universit\'e de Strasbourg, CNRS, Institut de Physique et Chimie des Mat\'eriaux de Strasbourg (IPCMS), UMR 7504, F-67000 Strasbourg, France}

\author{St\'ephane Berciaud}
\email{stephane.berciaud@ipcms.unistra.fr}
\affiliation{Universit\'e de Strasbourg, CNRS, Institut de Physique et Chimie des Mat\'eriaux de Strasbourg (IPCMS), UMR 7504, F-67000 Strasbourg, France}


\begin{abstract}
Van der Waals heterostuctures, made from stacks of two-dimensional materials, exhibit unique light-matter interactions and are promising for novel optoelectronic devices. The performance of such devices is governed by near-field coupling through, e.g., interlayer charge and/or energy transfer. New concepts and experimental methodologies are needed to properly describe two-dimensional heterointerfaces. Here, we report an original study of interlayer charge and energy transfer in atomically thin metal (graphene)/semiconductor (transition metal dichalcogenide (TMD, here MoSe$_2$)) heterostructures using a combination of micro-photoluminescence and Raman scattering spectroscopies. The photoluminescence intensity in graphene/MoSe$_2$ is quenched by more than two orders of magnitude and rises linearly with the photon flux, demonstrating a drastically shortened ($\sim 1~\tr{ps}$) room temperature MoSe$_2$ exciton lifetime. Key complementary insights are provided from a comprehensive analysis of the graphene and MoSe$_2$ Raman modes, which reveals net photoinduced electron transfer from MoSe$_2$ to graphene and hole accumulation in MoSe$_2$. Remarkably, the steady state Fermi energy of graphene saturates at $290\pm 15~\tr{meV}$ above the Dirac point. This reproducible behavior is observed both in ambient air and in vacuum and is discussed in terms of intrinsic factors (i.e., band offsets) and environmental effects. In this saturation regime, balanced photoinduced flows of electrons and holes may transfer to graphene, a mechanism that effectively leads to energy transfer.  Using a broad range of photon fluxes and diverse environmental conditions, we find that the presence of net photoinduced charge transfer has no measurable impact on the near-unity photoluminescence quenching efficiency in graphene/MoSe$_2$. This absence of correlation strongly suggests that energy transfer to graphene (either in the form of electron exchange or dipole-dipole interaction) is the dominant interlayer coupling mechanism between atomically-thin TMDs and graphene.

\end{abstract}

\maketitle

\section{Introduction}

Charge and energy transfer (CT, ET) play a prominent role in atomic, molecular and nanoscale systems. On the one hand, F\"ortser-type energy transfer~\cite{Forster1948}, mediated by relatively long-range (up to several nm) near-field dipole-dipole coupling is an essential step in photosynthesis~\cite{VANGRONDELLE1994} and is now engineered in a variety of light-harvesting devices and distance sensors~\cite{Guzelturk2016,Govorov2016}. Charge transfer, on the other hand is a much shorter range process ($\sim 1~\tr{nm}$) that plays a key role in a number of molecular and solid-state systems and is at the origin of the operation of photodetectors and solar cells~\cite{May2008,Sze2006}. In the limit of orbital overlap between donor and acceptor systems, electron exchange, resulting in no net charge transfer and also known as Dexter-type energy transfer~\cite{Dexter1953}, may occur. The efficiencies of CT and ET depend very sensitively on the donor-acceptor distance, on the energy levels (or bands) offsets, and on the local dielectric and electrostatic environment. CT and ET processes may have beneficial or detrimental impact on the performance of optoelectronic devices and therefore deserve fundamental investigations. 

In this context, two-dimensional materials (2DM, such as graphene, boron nitride, transition metal dichalcogenides (TMDs), black phosphorus,\dots) provide an extraordinary toolkit to investigate novel regimes of CT/ET. Indeed the very diverse and complementary physical properties of 2DM can be tailored and controlled at the single-layer level, but also combined and possibly enhanced within so-called van der Waals heterostructures (vdWHs)~\cite{Geim2013,Novoselov2016,Mak2016}. VdWHs provide a new paradigm of clean, ultra smooth two-dimensional heterointerfaces~\cite{Haigh2012}. Since their van der Waals gap is only of a few \angstrom, band bending and depletion regions cannot develop in vdWH. As a result, well-established concepts borrowed from the physics of bulk or low-dimensional heterojunctions~\cite{Sze2006} must be adapted with great care when describing the optoelectronic response of vdWHs. In addition, the ultimate proximity between the atomically thin building blocks that compose a vdWH potentially allows ultra efficient CT and/or ET.

Among the vast library of 2DM, graphene~\cite{Castroneto2009} and atomically-thin semiconducting TMDs (with formula MX$_2$, with M = Mo, W and X = S, Se, Te)~\cite{Mak2010,Splendiani2010,Mak2016} have attracted particular interest for optoelectronic applications~\cite{Britnell2013,Yu2013,Roy2013b,Zhang2014,DeFazio2016,Massicotte2016,Mccreary2014,Shim2014,Pierucci2016b,He2014b,Henck2016}. Indeed, graphene (Gr) may act as a highly tunable transparent electrode, endowed with exceptional physical properties~\cite{Koppens2014,Mak2012,Tielrooij2015}, while monolayer TMDs  are direct bandgap semiconductors with unusually strong light-matter interactions and excitonic effects~\cite{Wang2017,Xia2014,Mak2016}, as well as unique spin, valley and optoelectronic properties~\cite{Xia2014,Mak2016,Schaibley2016}. Photodetectors based on graphene and TMDs display high photoresponsivity and photogain~\cite{Britnell2013,Yu2013,Roy2013b,Zhang2014,DeFazio2016}, down to picosecond timescales~\cite{Massicotte2016}. The photophysics of Gr/TMD vdWHs is governed by near-field interlayer CT and/or ET (ICT, IET).  In the related and most studied case of TMD/TMD heterojunctions with type II band alignment, sub-picosecond ICT and subsequent interlayer exciton formation is thought to be the dominant coupling mechanism~\cite{Ceballos2014,Hong2014,Fang2014,Lee2014a,Rivera2015,Schaibley2016}. However, recent photoluminescence excitation spectroscopy studies in MoSe$_2$/WS$_2$ vdWH have suggested that IET may be at least as efficient as ICT~\cite{Kozawa2016}.

	In contrast, fundamental studies of IET and ICT remain scarce in Gr/TMD vdWH. Photoinduced ICT has been observed in Gr/MoS$_2$ photodetectors~\cite{Zhang2014}. Recent transient absorption studies have evidenced fast interlayer coupling in Gr/WS$_2$ vdWHs and tentatively assigned it to photoinduced ICT~\cite{He2014b}. Yet, such studies were mostly performed under ambient conditions and the share of environmental effects needs to be assessed. Importantly, in Ref.~\onlinecite{Massicotte2016}, the internal quantum efficiency of Gr/TMD photodetectors degrades when the active TMD layer is thinned down to the monolayer limit, possibly due to efficient IET to graphene.
	Overall, IET has been surprisingly overlooked in vdWH, whereas  related studies in hybrid heterostructures composed of nanoscale emitters (molecules, quantum dots, quantum wells,\dots) interfaced with carbon nanotubes~\cite{Roquelet2010}, graphene~\cite{Chen2010,Gaudreau2013,Tisler2013,Federspiel2015}, TMDs~\cite{Prins2014,Raja2016} have consistently demonstrated highly efficient F\"orster-type ET. 

Unraveling the relative efficiencies of ICT and IET in vdWH is a timely challenge for optoelectronics. For this purpose, optical spectroscopy offer minimally invasive and spatially-resolved tools. First exciton dynamics and interlayer coupling can be probed with great sensitivity using micro-photoluminescence (PL) spectroscopy~\cite{Mak2016,Schaibley2016}. Second, micro-Raman scattering spectroscopy allows quantitative measurements of doping and charge transfer as it has been demonstrated in graphene~\cite{Yan2007,Pisana2007,Das2008,Froehlicher2015a,Ryu2010} and in MoS$_2$~\cite{Chakraborty2012,Miller2015}, but not yet in vdWHs.

In this paper, using an original combination of PL and Raman spectroscopies,  we are able to disentangle contributions from ICT and IET in model vdWHs made of single-layer graphene stacked onto single-layer molybdenum diselenide (MoSe$_2$) (hereafter denoted Gr/MoSe$_2$) \textit{in the absence of any externally applied electric field}. While highly efficient exciton-exciton annihilation and subsequent saturation of the PL intensity is -- as expected -- observed in bare MoSe$_2$ as the incident photon flux increases, the PL in Gr/MoSe$_2$ is massively quenched and its intensity rises linearly with the photon flux, demonstrating a drastically shortened room-temperature exciton lifetime in MoSe$_2$. Key complementary insights are provided from an comprehensive analysis of the graphene and MoSe$_2$ Raman modes, which reveals net photoinduced electron transfer from MoSe$_2$ to graphene and hole accumulation in MoSe$_2$. Remarkably,  the steady state Fermi energy of graphene saturates at $290\pm 15~\tr{meV}$ above the Dirac point. In this saturation regime, balanced flows of electrons and holes transfer to graphene, resulting in no net photoinduced charge transfer. This reproducible behavior is  observed both in ambient air and in vacuum and is discussed in terms of intrinsic factors (i.e., band offsets) and extrinsic effects associated with native doping and charge trapping.  Using a broad range of photon fluxes and diverse environmental conditions, we find that the existence of net photoinduced charge transfer has no measurable impact on the near-unity photoluminescence quenching efficiency in graphene/MoSe$_2$. This absence of correlation strongly suggests that energy transfer to graphene (either in the form of Dexter or F\"orster processes) is the dominant interlayer coupling mechanism between atomically-thin TMDs and graphene. Our results provide a better understanding of the atomically thin two-dimensional metal-semiconductor (i.e., Schottky) junction, an ubiquitous building block in emerging optoelectronic devices, and will serve as a guide to engineer charge carrier and exciton transport in two-dimensional materials.

\begin{figure*}[!tbh]
\begin{center}
\includegraphics[width=0.9\linewidth]{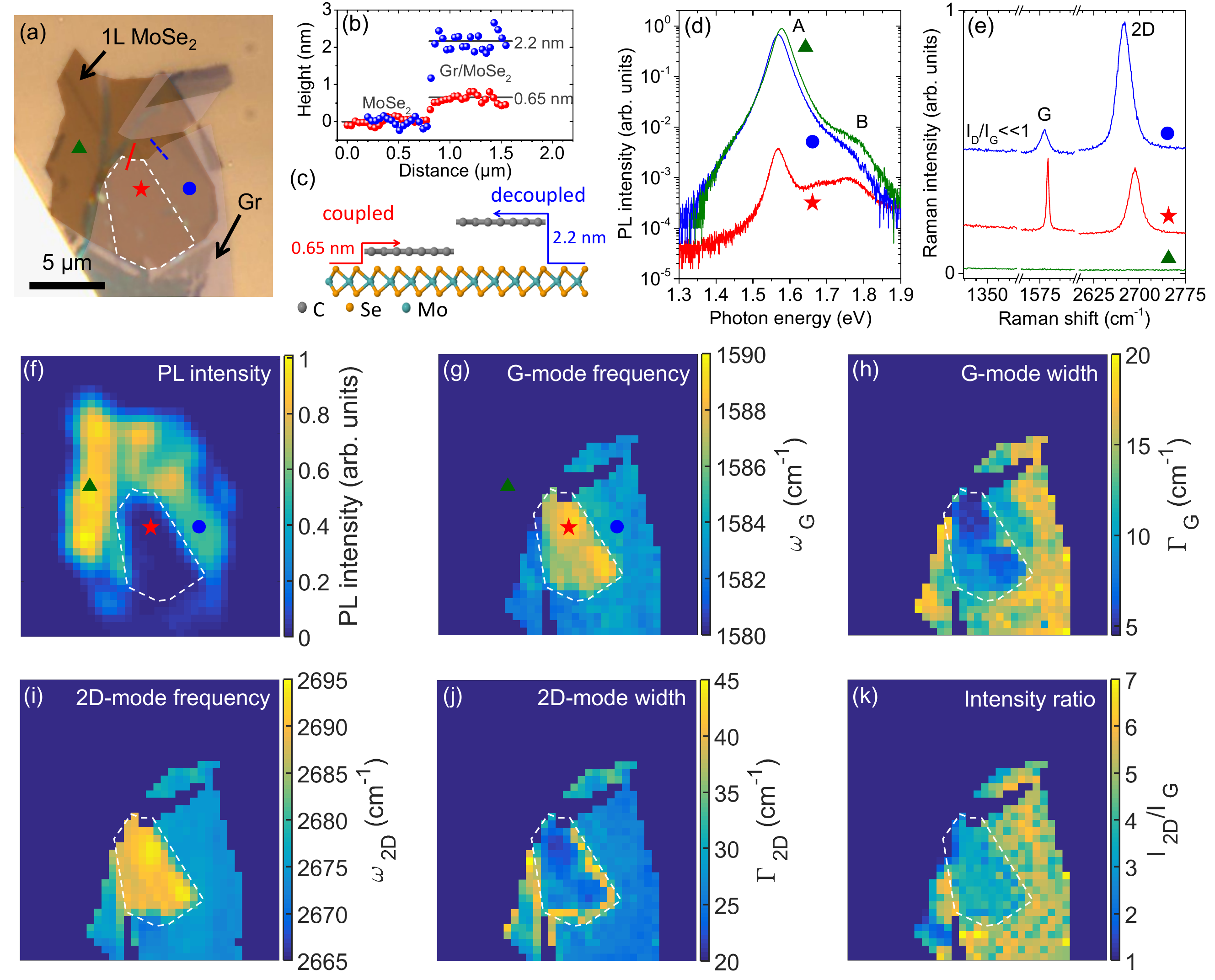}
\caption{(a) Optical image of a single-layer graphene/single-layer MoSe$_2$ van der Waals heterostructure deposited onto a Si/SiO$_2$ substrate (sample S$_1$). The coupled part of the heterostructure is represented by the white dashed contour. (b) Height profiles, measured by atomic force microscopy, along the dashed blue and red lines drawn in (a). (c) Schematic of the heterostructure showing the coupled and decoupled regions. Photoluminescence (d) and Raman scattering (e) spectra recorded on the three spots shown in (f) and (g), respectively. In (d) and (e), the spectra are plotted  with the same color as the symbols in (f) and (g), respectively. (f) MoSe$_2$ photoluminescence intensity map. (g-k) Hyperspectral Raman maps  of the (g) G-mode frequency $\omega_\tr{G}$, (h) G-mode FWHM $\Gamma_\tr{G}$, (i) 2D-mode frequency $\omega_\tr{2D}$, (j) 2D-mode FWHM $\Gamma_\tr{2D}$, and (k) ratio between the integrated intensities of the 2D- and G-mode features ($I_\tr{2D}/I_\tr{G}$) measured on single-layer graphene. All maps have the same scale as in (a) and were recorded in ambient air at a laser photon energy of 2.33~eV, with an incident photon flux $\Phi_\tr{ph}=2\times10^{19}~\tr{cm}^{-2}~\tr{s}^{-1}$ and $\Phi_\tr{ph}=2\times10^{22}~\tr{cm}^{-2}~\tr{s}^{-1}$ for photoluminescence and Raman measurements, respectively.}
\label{Fig1}
\end{center}
\end{figure*}

\section{Characterization of the G\lowercase{r}/M\lowercase{o}S\lowercase{e}$_2$ heterostructure}
\label{Carac}
 Figure~\ref{Fig1}(a) shows an optical image of a Gr/MoSe$_2$ vdWH (Sample S$_1$) deposited onto a Si/SiO$_2$ substrate. From AFM measurements (see Supplemental Material~\cite{SMnote}, Fig.~S1), we can distinguish a region of the heterostucture (highlighted with a white dashed contour in Fig.~\ref{Fig1}(a)), where the two layers are well coupled, as evidenced by the small surface roughness~\cite{Novoselov2016} and the small height difference of approximately $0.65~\tr{nm}$ between the surface of MoSe$_2$ and Gr (see Fig.~\ref{Fig1}(b)). Outside this region, the interface shows sub-micrometer size {``pockets'' and an average step of  $\sim 2-3~\tr{nm}$ (see Fig.~\ref{Fig1}(b)) between MoSe$_2$ and Gr. Hereafter, the former and the latter are referred to \textit{coupled} and \textit{decoupled} Gr/MoSe$_2$, respectively (see Fig\ref{Fig1}(c) and Supplemental Material~\cite{SMnote}, Fig.~S1).

Typical photoluminescence (PL) and Raman spectra from three different points of the sample are shown in Fig.~\ref{Fig1}(d) and Fig.~\ref{Fig1}(e), respectively. Unless otherwise noted, the samples were optically excited in the continuous wave regime using a single longitudinal mode, linearly polarized laser at a photon energy of $2.33~\tr{eV}$ well above the optical bandgap of MoSe$_2$.

Figure.~\ref{Fig1}(f-k) displays the hyperspectral maps of (f) the MoSe$_2$ PL intensity, (g-j) the frequencies ($\omega_\tr{G,2D}$) and full-widths at half maximum (FWHM, $\Gamma_\tr{G,2D}$) of the Raman G- and 2D-mode features~\cite{Ferrari2013}, and (k) of the ratio of their integrated intensities ($I_\tr{2D}/I_{\tr G}$). Note that no defect-induced D-mode feature~\cite{Ferrari2013} (expected around $1350~\tr{cm}^{-1}$) emerges from the background showing the very good quality of our sample. All hyperspectral maps allow to distinctively identify the coupled and decoupled Gr/MoSe$_2$ regions and confirm the trends observed on selected points.

The PL spectra in Fig.~\ref{Fig1}(d) are characteristic of single-layer MoSe$_2$ with the A and B excitons~\cite{Li2015} near 1.57~eV and 1.75~eV, respectively.  Remarkably, the MoSe$_2$ PL intensity  is $\sim 300$ times smaller on coupled Gr/MoSe$_2$ than on MoSe$_2$/SiO$_2$, while it is only reduced by a modest factor of $\sim 2$ on decoupled Gr/MoSe$_2$ (Fig.~\ref{Fig1}(d,f). Such massive PL quenching, also observed for other Gr/TMD vdWHs~\cite{He2014b,Massicotte2016} demonstrates strong interlayer coupling and suggest a much reduced exciton lifetime. 
\begin{figure*}[!tbh]
\begin{center}
\includegraphics[width=0.7\linewidth]{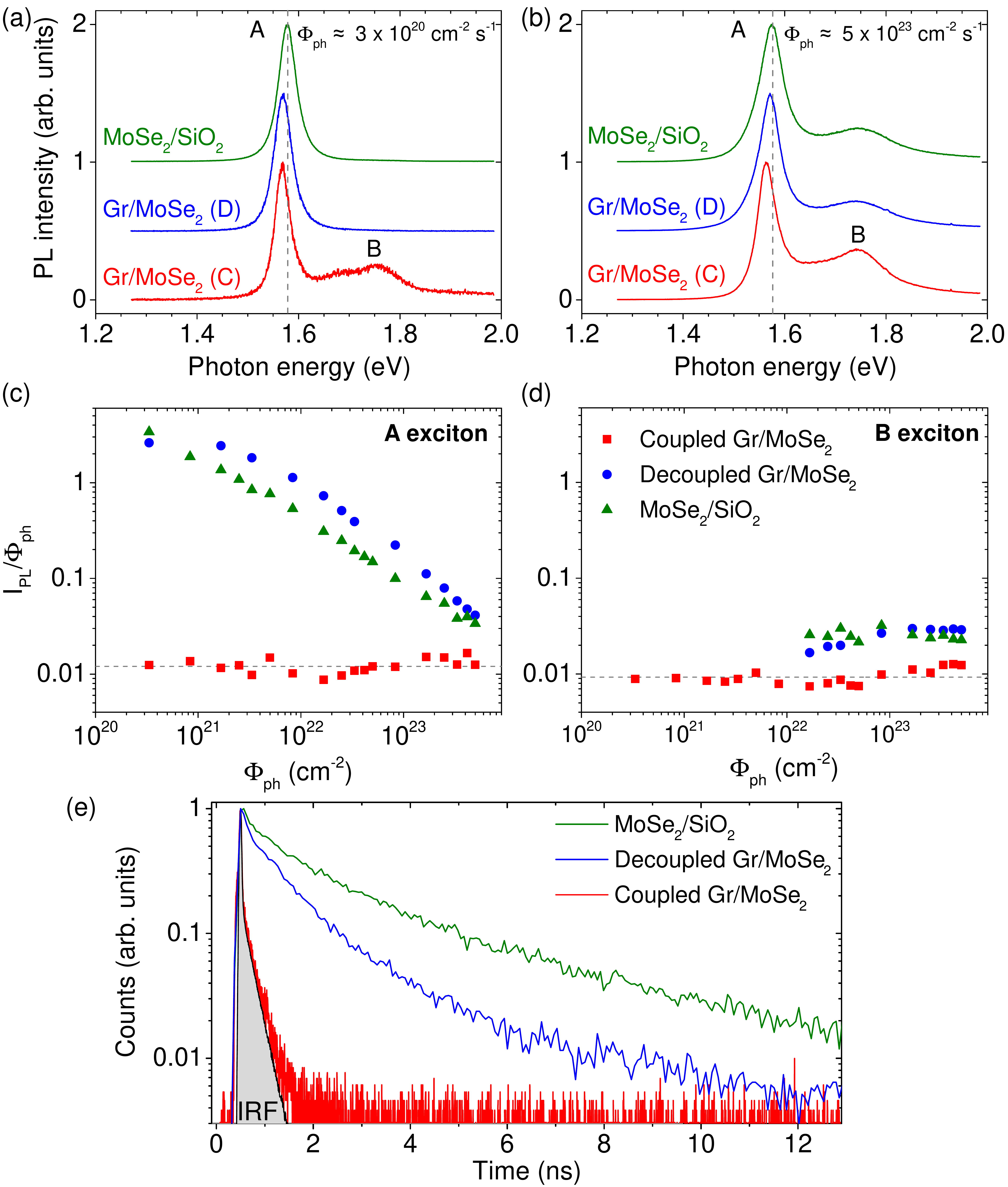}
\caption{Normalized photoluminescence spectra of MoSe$_2$ recorded in the cw regime at a laser photon energy of 2.33~eV, with an incident photon flux (a) $\Phi_\tr{ph}=3\times10^{20}~\tr{cm}^{-2}~\tr{s}^{-1}$ and (b) $\Phi_\tr{ph}=5\times10^{23}~\tr{cm}^{-2}~\tr{s}^{-1}$  for MoSe$_2$/SiO$_2$, decoupled (D) and coupled (C) Gr/MoSe$_2$. The A and B excitons are labeled and the position of the A exction in MoSe$_2$/SiO$_2$ is indicated by a gray vertical dashed line. (b) Integrated photoluminescence intensity of the (c) A and (d) B exciton normalized by $\Phi_\tr{ph}$ as a function of $\Phi_\tr{ph}$. The gray dashed line is a guide to the eye. The error bars are smaller than symbol size. (e) Photoluminescence decays recorded using a pulsed laser at 1.96~eV with a fluence of $\approx 2.2\times10^{11}~\tr{cm}^{-2}$ per pulse. The gray area corresponds to the instrument response function (IRF). All measurements were performed in ambient air.}
\label{FigPL}
\end{center}
\end{figure*}  

As shown in Fig.~\ref{Fig1}(e),(g-k), interlayer coupling also dramatically affects the Raman response of graphene. Indeed, on coupled Gr/MoSe$_2$, the G-mode feature upshifts, gets narrower, and the $I_\tr{2D}/I_{\tr G}$ ratio decreases (Fig.~\ref{Fig1}(k)) with respect to reference measurements on the neighboring pristine graphene deposited on SiO$_2$ (Gr/SiO$_2$) and decoupled Gr/MoSe$_2$ regions (Fig.~\ref{Fig1}(e,k)). These observations are robust evidence of an increased charge carrier concentration in graphene~\cite{Das2008,Froehlicher2015a}. Surprisingly, we observe an upshift of the 2D-mode frequency on coupled Gr/MoSe$_2$ (Fig.~\ref{Fig1}(i)), which is too high to be solely induced by doping or strain~\cite{Lee2012,Froehlicher2015a,Metten2014}, and seems qualitatively similar to previous reports on graphene deposited on thick boron nitride terraces~\cite{Ahn2013,Forster2013} and monolayer MoS$_2$ grown on graphene~\cite{Mccreary2014}. Possible origins for this upshift are discussed in the Supplemental Material~\cite{SMnote} (Fig.~S15). 

In the following we quantitatively investigate exciton dynamics (Sec.~\ref{SecPL}) and interlayer charge transfer (Sec.~\ref{SecICT}) .

\section{Exciton dynamics in G\lowercase{r}/M\lowercase{o}S\lowercase{e}$_2$}
\label{SecPL}

Although PL quenching has been reported in previous studies of Gr/TMD heterostructures (see Figure 1 in Ref.~\onlinecite{He2014} and Supplementary Figure 6 in Ref.~\cite{Massicotte2016}), quantitative analysis of PL quenching and its interpretation in terms of IET and ICT have not been reported thus far.

Figure~\ref{FigPL}(a,b) displays the normalized PL spectra of MoSe$_2$ recorded on MoSe$_2$/SiO$_2$, decoupled and coupled Gr/MoSe$_2$ at low and high incident photon flux $\Phi_{\tr{ph}}$. The A exciton PL feature of coupled Gr/MoSe$_2$/SiO$_2$ is marginally redshifted (by $\approx 10~\tr{meV}$) with respect to that of air/MoSe$_2$/SiO$_2$, irrespective of $\Phi_{\tr{ph}}$. Considering the drastically different dielectric environments, such a surprisingly small reduction of the \textit{optical} bandgap is assigned to the near-perfect compensation of the reductions of electronic bandgap and exciton binding energy in graphene-capped MoSe$_2$\cite{Ugeda2014,Stier2016,Raja2017}. The lineshapes of the A exciton features are quite similar, except for a small but reproducible narrowing of the A-exciton linewidth in coupled Gr/MoSe$_2$. Similar narrowing has recently been observed in TMD layers fully encapsulated in boron nitride~\cite{Cadiz2017,Ajayi2017} and likely results from a reduction of inhomogeneous broadening and pure dephasing in graphene-capped TMD samples.

\begin{figure*}[!t]
\begin{center}
\includegraphics[width=0.9\linewidth]{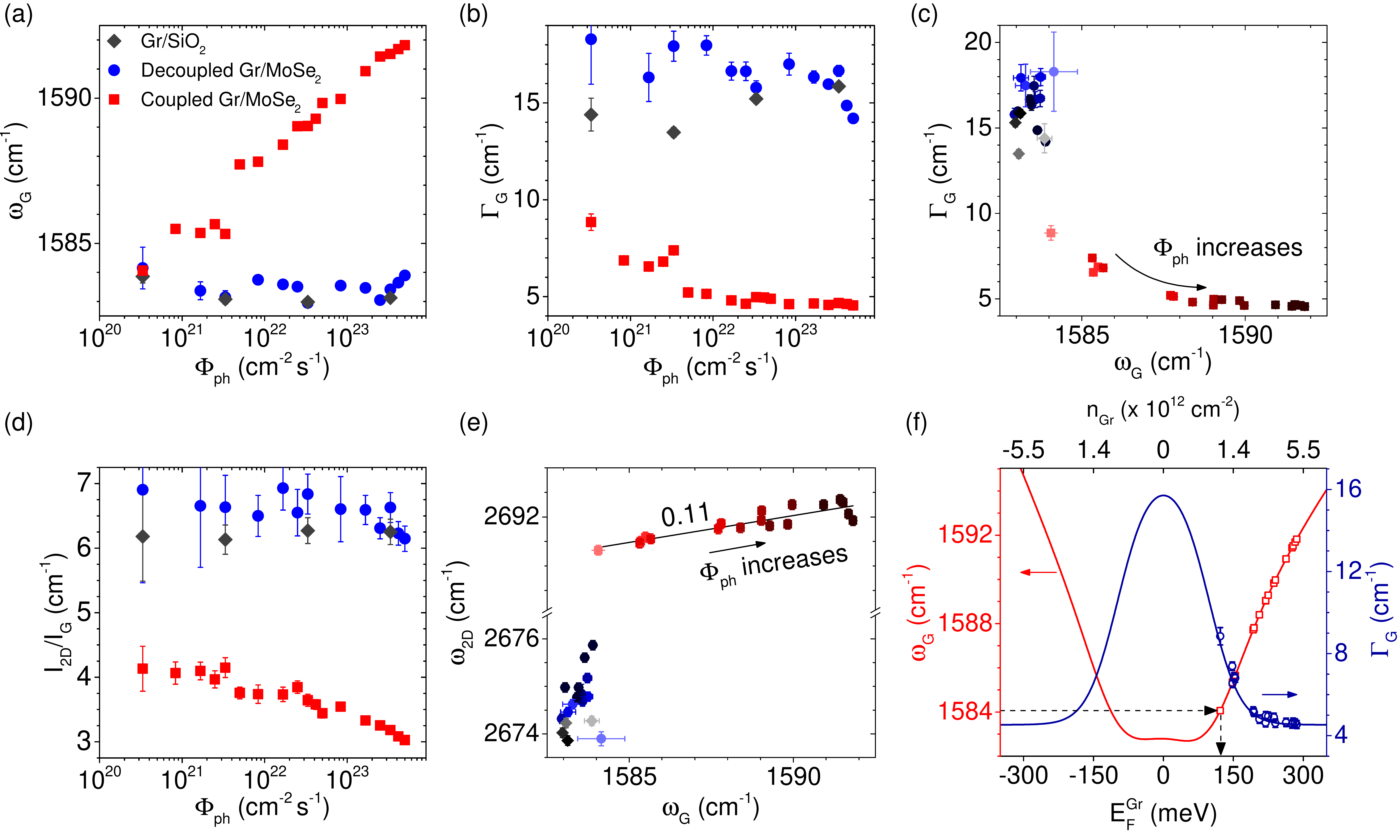}
\caption{(a) Raman G-mode frequency $\omega_\tr{G}$ and (b) full-width at half maximum $\Gamma_\tr{G}$ recorded at a laser photon energy of 2.33~eV as a function of the incident photon flux $\Phi_\tr{ph}$ for coupled (red squares) and decoupled (blue circles) Gr/MoSe$_2$ and for Gr/SiO$_2$ (gray diamonds) (see Figure~\ref{Fig1}). (c) Correlation between $\Gamma_\tr{G}$ and $\omega_\tr{G}$ under increasing $\Phi_\tr{ph}$. (d) ratio between the integrated intensity of the 2D-mode feature and that of the G-mode feature $I_\tr{2D}/I_\tr{G}$ as a function of $\Phi_\tr{ph}$. (e) Correlation between $\omega_\tr{2D}$ and $\omega_\tr{G}$ under increasing $\Phi_\tr{ph}$. The color of the symbols in (c) and (e) gets darker with increasing $\Phi_\tr{ph}$. (f) $\omega_\tr{G}$ (red squares, left axis) and $\Gamma_\tr{G}$ (blue circles, right axis) as a function of the graphene Fermi energy $E^\tr{Gr}_\tr{F}$  and doping level $n_\tr{Gr}$. The solid lines are theoretical calculations~\cite{Pisana2007,Froehlicher2015a}. All measurements were performed in ambient air.}
\label{FigICT}
\end{center}
\end{figure*}

The integrated PL intensities of the A and B exciton features (denoted $I_\tr{PL}^\tr{A,B}$) normalized by $\Phi_\tr{ph}$, are plotted as a function of $\Phi_\tr{ph}$ in Figure~\ref{FigPL}(c) and (d), respectively. For MoSe$_2$/SiO$_2$ and decoupled Gr/MoSe$_2$, $I_\tr{PL}^{\tr A}/\Phi_{\tr{ph}}$ drops abruptly as $\Phi_{\tr{ph}}$ augments due to highly efficient exciton-exciton annihilation (EEA), as previously evidenced in TMD monolayers~\cite{Kumar2014,Mouri2014}. In the case of coupled Gr/MoSe$_2$, $I_\tr{PL}^{\tr A}/\Phi_{\tr{ph}}$  remains constant, within experimental accuracy, up to  $\Phi_{\tr{ph}}\sim 10^{24}~\tr{cm}^{-2}~\tr{s}^{-1}$. As a result, while $I_\tr{PL}^{\tr A}$ is about 300 times weaker on coupled Gr/MoSe$_2$ than on bare MoSe$_2$ at $\Phi_{\tr{ph}}=3\times 10^{20}~\tr{cm}^{-2}~\tr{s}^{-1}$, this quenching factor reduces down to $\sim 3$ at $\Phi_{\tr{ph}}=  6 \times 10^{23}~\tr{cm}^{-2}~\tr{s}^{-1}$. Strong PL quenching together with the linear  scaling of $I_\tr{PL}^{\tr A}$ with $\Phi_{\tr{ph}}$  demonstrate that interlayer coupling between graphene and MoSe$_2$ opens up non-radiative decay channel for A excitons, that dramatically  reduces of the A exciton lifetime and is sufficiently fast to bypass EEA. Very similar PL quenching and exciton dynamics have been observed in other Gr/MoSe$_2$/SiO$_2$ samples (see Supplemental Material~\cite{SMnote}, Fig. S11-S13) as well as in Gr/WS$_2$/SiO$_2$ (see Supplemental Material~\cite{SMnote}, Fig. S14) and Gr/WSe$_2$/SiO$_2$ (data not shown).
The shortening of the A exciton lifetime is further substantiated by the analysis of the \textit{hot} luminescence from the B exciton (note that our samples are photoexcited at 2.33~eV, i.e., well-above the B exciton in MoSe$_2$). In bare MoSe$_2$ and decoupled Gr/MoSe$_2$, $I_\tr{PL}^{\tr A}\gg I_\tr{PL}^{\tr B}$, whereas $I_\tr{PL}^\tr{B}\sim I_\tr{PL}^{\tr A}$ in coupled Gr/MoSe$_2$. Interestingly, $I_\tr{PL}^{\tr B}$ is very similar in the three cases and scales linearly with $\Phi_\tr{ph}$ (see Fig.~\ref{FigPL}(d)). These observations suggest (i) that interlayer coupling does not significantly affect exciton formation and exciton decay until a population of A excitons is formed, and (ii) that the A exciton lifetime in Gr/MoSe$_2$ is not appreciably longer than the $\tr{B}\rightarrow \tr{A}$ decay time. The latter is typically in the subpicosecond range~\cite{Shi2013} in atomically thin TMDs, and provides a lower bound for the A exciton lifetime in Gr/MoSe$_2$. Additional insights are provided by time-resolved photoluminescence measurements recorded in ambient conditions (see Fig.~\ref{FigPL}(e)). Bare MoSe$_2$ and decoupled Gr/MoSe$_2$ display non-monoexponential decays~\cite{Robert2016} with average exciton lifetime of $\sim 1~\tr{ns}$. As anticipated, the PL decay of Gr/MoSe$_2$ is too fast to be resolved using our experimental apparatus, confirming that the A exciton lifetime is significantly shorter that our time-resolution of $\sim 20~\tr{ps}$. Using the estimated decay time of bare MoSe$_2$ and a typical quenching factor of $\sim 300$ (i.e., a quenching efficiency of $\sim 99.7\%$) in the low fluence limit, we can reckon a conservative upper bound of a few ps for the exciton lifetime in coupled Gr/MoSe$_2$.


\begin{figure*}[!t]
\begin{center}
\includegraphics[width=0.7\linewidth]{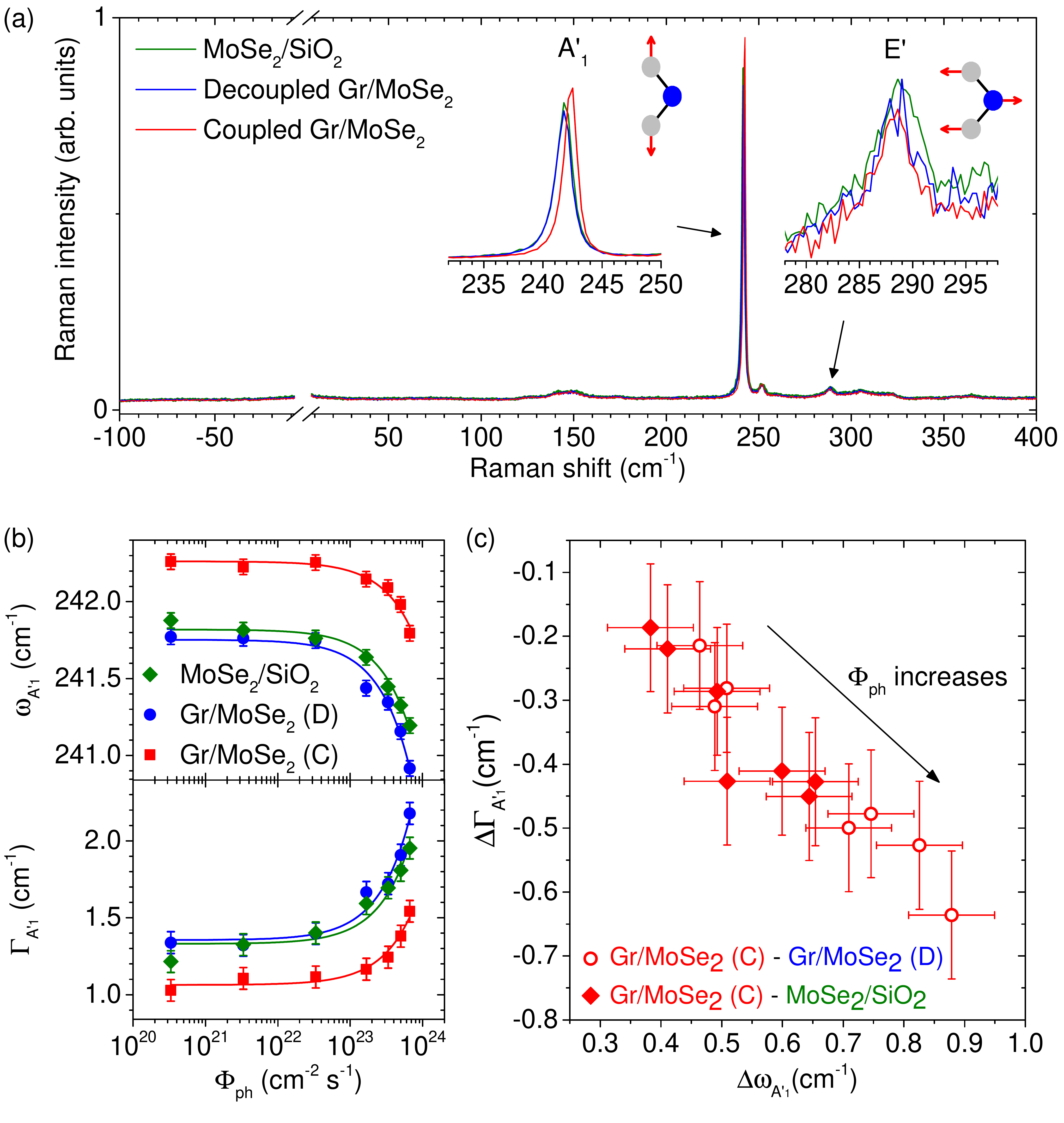}
\caption{(a) Raman spectra of MoSe$_2$ recorded at 2.33~eV, with $\Phi_\tr{ph}\approx 3.3\times10^{22}~\tr{cm}^{-2}~\tr{s}^{-1}$ for MoSe$_2$/SiO$_2$ (green), coupled (C, in red) and decoupled (D, in blue) Gr/MoSe$_2$. The insets show a close-up on the two one-phonon Raman-active modes in single-layer MoSe$_2$, namely the $A^{\prime}_1$ and $E^{\prime}$ modes. The corresponding atomic displacements are sketched. (b) Extracted  frequency (top) and  FWHM (bottom) of the $A^{\prime}_1$-mode feature as a function of incident photon flux $\Phi_\tr{ph}$ for MoSe$_2$/SiO$_2$ (green diamonds), coupled (red squares) and decoupled (blue circles) Gr/MoSe$_2$. The solid lines are linear fits. (c) Difference between the $A^{\prime}_1$-mode FWHMs $\Delta\Gamma_{A^{\prime}_1}$ as a function of the difference between the $A^{\prime}_1$-mode frequencies $\Delta\omega_{A^{\prime}_1}$ for coupled and decoupled Gr/MoSe$_2$ (open circles), and for coupled Gr/MoSe$_2$ and MoSe$_2$/SiO$_2$ (filled diamonds). All measurements were performed in ambient air.}
\label{FigMoSe2}
\end{center}
\end{figure*}

\section{Interlayer charge transfer}
\label{SecICT}
\subsection{Net photoinduced electron transfer to graphene}
\label{SecICTGr}

The fast MoSe$_2$ exciton decay in Gr/MoSe$_2$ heterostructures may arise from a combination of ICT and IET processes. In this section, we introduce an original Raman-based readout of the steady state charge carrier density in both materials.

Fig.~\ref{FigICT} shows the evolution of $\omega_\tr{G,2D}$, $\Gamma_\tr{G,2D}$ and $I_\tr{2D}/I_{\tr G}$ measured in sample S$_1$ as a function of $\Phi_\tr{ph}$, in ambient air. The corresponding spectra are shown in the Supplemental Material~\cite{SMnote} (Fig.~S2). First, for Gr/SiO$_2$ and decoupled Gr/MoSe$_2$, $\omega_\tr{G}\approx 1583~\tr{cm}^{-1}$, $\Gamma_\tr{G}\approx 16~\tr{cm}^{-1}$, $\omega_\tr{2D}\approx 2674~\tr{cm}^{-1}$ and $I_\tr{2D}/I_\tr{G}\approx 6.5$ do not show any appreciable variation as $\Phi_\tr{ph}$ augments. These values correspond to very weakly doped graphene ($\abs{n_{\tr Gr}}\sim 10^{11}~\tr{cm}^{-2}$ or $\abs{E^{\tr Gr}_\tr{F}}\lesssim100~\tr{meV}$)~\cite{Das2008,Berciaud2009,Froehlicher2015a}. In addition, the absence of measurable phonon softening at high $\Phi_\tr{ph}$, indicates that the laser-induced temperature rise remains below $\sim 100~\tr{K}$~\cite{Balandin2011}.

Second, for coupled Gr/MoSe$_2$, $\omega_{\tr G}$ distinctly rises  as $\Phi_\tr{ph}$ increases, whereas $\Gamma_{\tr G}$ decreases (see Figs.~\ref{FigICT}(a)-(c)). Additionally, $I_\tr{2D}/I_\tr{G}$ (Figs.~\ref{FigICT}(e)) drops significantly. These spectroscopic features provide strong evidence for \textit{net photoinduced} ICT from MoSe$_2$ to graphene~\cite{Das2008,Froehlicher2015a}. We can now identify the sign of the net transferred charge flow using the correlation between $\omega_\tr{2D}$ and $\omega_{\tr G}$  in Fig.~\ref{FigICT}(f) as in Ref.~\onlinecite{Lee2012,Froehlicher2015a}. As $\Phi_\tr{ph}$ increases, the data for Gr/SiO$_2$ and decoupled Gr/MoSe$_2$ show no clear correlations. In contrast, on coupled Gr/MoSe$_2$ $\omega_\tr{2D}$ and $\omega_{\tr G}$ display a linear correlation with a slope of $\approx 0.11$, a value that clearly points towards photoinduced electron doping in graphene~\cite{Froehlicher2015a}. 

Using well-established theoretical modelling of electron-phonon coupling in doped graphene~\cite{Ando2006,Pisana2007}, we quantitatively determine the Fermi energy of graphene relative to the Dirac point $E^\tr{Gr}_\tr{F}$ or equivalently its doping level $n_\tr{Gr}$. The values of $E^\tr{Gr}_\tr{F}$ and $n_\tr{Gr}$  extracted from a global fitting procedure (see Ref.~\onlinecite{Froehlicher2015a}) are plotted in Fig.~\ref{FigICT}(f). As further discussed in Sec.~\ref{AirVac}, $E^\tr{Gr}_\tr{F}$ ($n_\tr{Gr}$) saturates as $\Phi_{\tr{ph}}$ increases and reaches up to $\approx 280~\tr{meV}$ ($\approx 5\times 10^{12}~\tr{cm}^{-2}$).

\subsection{Hole accumulation in MoSe$_2$}

Net electron transfer to graphene naturally implies hole accumulation in MoSe$_2$. Depending on the initial doping of MoSe$_2$, photoinduced hole accumulation in MoSe$_2$ should allow or impede the formation of charged excitons (trions). However, at room temperature, trions in MoSe$2$ are not stable enough~\cite{Ross2013} to allow the observation of trion emission in our PL spectra. However, as in Sec.~\ref{SecICTGr}, we can seek for fingerprints of ICT in the high-resolution Raman response of MoSe$_2$. 

Figure~\ref{FigMoSe2}(a) shows the MoSe$_2$ Raman spectra from MoSe$_2$/SiO$_2$, decoupled and coupled Gr/MoSe$_2$. In addition to several higher-order resonant features involving finite momentum phonons, one can identify the two Raman-active one-phonon modes in monolayer MoSe$_2$ with $A^{\prime}_1$ symmetry (near $242~\tr{cm}^{-1}$) and $E^{\prime}$ symmetry (near $289~\tr{cm}^{-1}) $~\cite{Soubelet2016,Zhang2015b}. The faint $E^{\prime}$ mode-feature is slightly downshifted on coupled Gr/MoSe$_2$, as compared to MoSe$_2$/SiO$_2$. The prominent $A^{\prime}_1$-mode feature is much similar for MoSe$_2$/SiO$_2$ and decoupled Gr/MoSe$_2$, but distinctively blueshifts (by $\approx 0.5~\tr{cm}^{-1}$) and gets narrower (by $\approx 20 \%$) for coupled Gr/MoSe$_2$ (see Fig.~\ref{FigMoSe2}(b)).

As in the case of graphene, changes in the Raman spectra can tentatively be assigned to doping, with possible spurious contributions from native strain, laser-induced heating, as well as van der Waals coupling~\cite{Zhou2014} and surface effects~\cite{Luo2013,Froehlicher2015b}\footnote{The observed upshift may in part stem from van der Waals coupling between the graphene and MoSe$_2$monolayers~\cite{Zhang2015b,Zhou2014,Luo2013,Froehlicher2015b} (similarly to the case of TMD bilayers), as well as from surface effects~\cite{Luo2013,Froehlicher2015b}, i.e. in the present case, slightly larger force constants between Mo and Se atoms in Gr/MoSe$_2$/SiO$_2$ than in air/MoSe$_2$/SiO$_2$. However both kinds of effects would not lead to the significant narrowing of the $A^{\prime}_1$ feature observed in Gr/MoSe$_2$/SiO$_2$ and cannot account for the differential effects shown in Fig.~\ref{FigMoSe2}c.}. Interestingly, recent Raman studies in MoS$_2$ monolayers have demonstrated that the $A^{\prime}_1$-mode phonon undergoes modest doping-induced phonon renormalization, namely a downshift and a broadening for increasing electron concentration whereas the $E^{\prime}$-mode phonon is largely insensitive to doping~\cite{Chakraborty2012,Miller2015}. Conversely, also in MoS$_2$, it was shown that under tensile (resp. compressive) strain the $E^{\prime}$-mode feature undergoes much larger shifts than the $A^{\prime}_1$-mode feature~\cite{Conley2013,Zhang2015b}. The $A^{\prime}_1$ and $E^{\prime}$ phonons may thus be used as probes of ICT and strain, respectively. Based on these considerations, the minute $E^{\prime}$ phonon softening observed irrespective of $\Phi_\tr{ph}$ in Gr/MoSe$_2$ relative to MoSe$_2$/SiO$_2$ suggests a slightly larger native tensile strain on Gr/MoSe$_2$, that has no impact whatsoever on ICT (see Supplemental Material~\cite{SMnote}, Fig.~S4-S8). More importantly, the upshifted and narrower $A^{\prime}_1$-mode feature consistently observed up to $\Phi_\tr{ph}\approx 6\times 10^{23}~\tr{cm}^{-2}~\tr{s}^{-1}$  in coupled Gr/MoSe$_2$ indicates a lower electron density in MoSe$_2$ than in decoupled Gr/MoSe$_2$ and MoSe$_2$/SiO$_2$.

However, on the three regions of the sample, the frequency and FWHM of the $A^{\prime}_1$-mode feature downshifts and increases linearly as $\Phi_\tr{ph}$ augments, respectively. Such trends counter-intuitively suggest photoinduced electron doping in MoSe$_2$. We tentatively assign the observed evolution of the $A^{\prime}_1$-mode feature to slight laser-induced temperature increase (estimated below $100~\tr{K}$~\cite{Late2014} at $\Phi_\tr{ph}\approx 6\times 10^{23}~\tr{cm}^{-2}~\tr{s}^{-1}$), possibly combined with related photogating effects involving the  presence of molecular  adsorbates and trapped charges both acting as electron acceptors and laser-assisted desorption of the latter~\cite{Miller2015}. Remarkably, as shown in Fig.~\ref{FigMoSe2}(c), the difference between the $A^{\prime}_1$ frequencies (FWHM) measured on coupled Gr/MoSe$_2$ and decoupled Gr/MoSe$_2$ or MoSe$_2$/SiO$_2$ monotonically increases (decreases) as $\Phi_\tr{ph}$ augments. These observations correspond to a \textit{net photoinduced hole doping} for MoSe$_2$ in coupled Gr/MoSe$_2$, \textit{relative to}  decoupled Gr/MoSe$_2$ and MoSe$_2$/SiO$_2$, consistently with the net  photoinduced electron transfer from MoSe$_2$ to graphene demonstrated in Fig.~\ref{FigICT}.

\section{Environmental effects}
\label{AirVac}

\begin{figure}[!t]
\begin{center}
\includegraphics[width=0.9\linewidth]{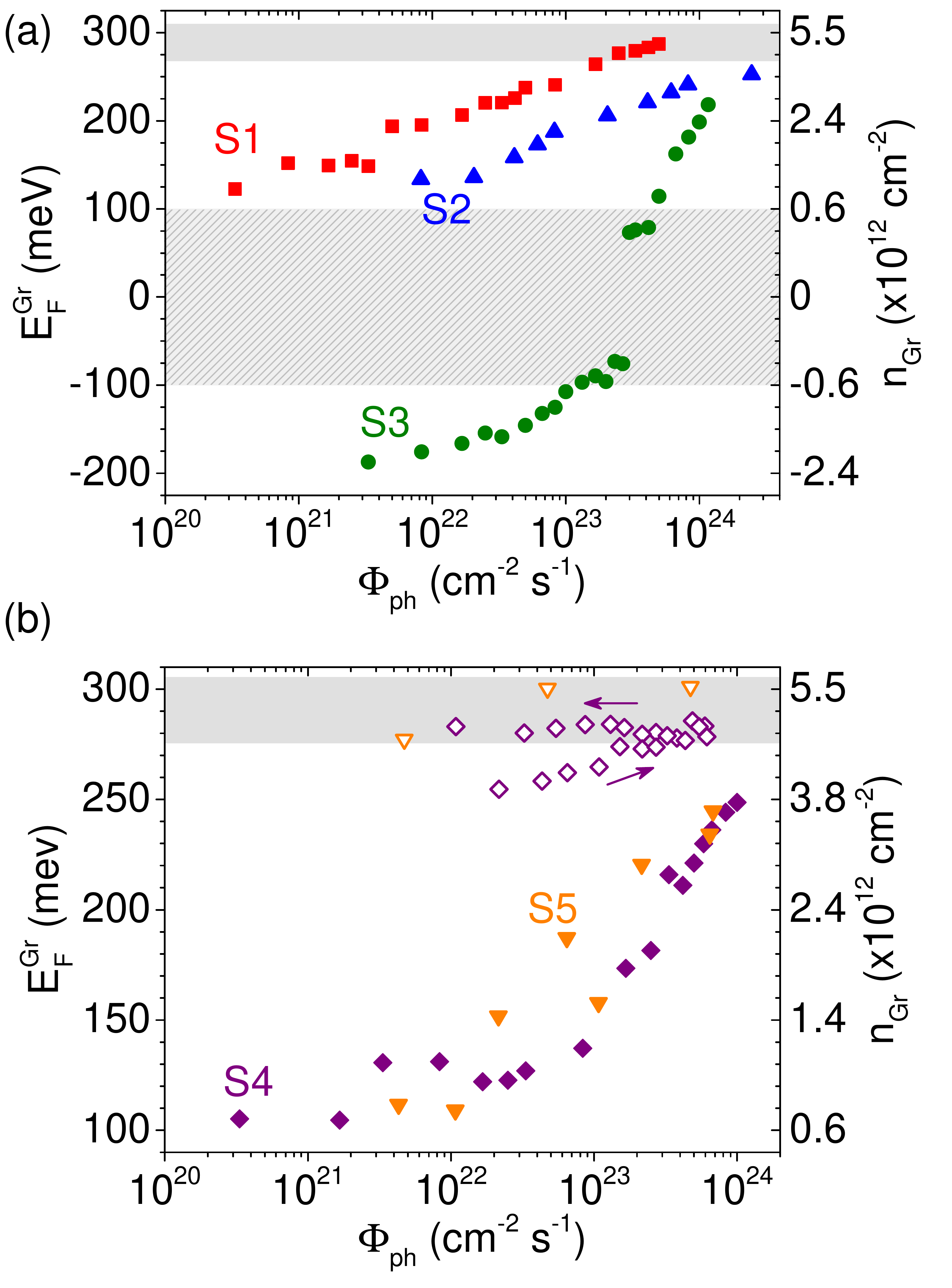}
\caption{ (a) Fermi energy $E^\tr{Gr}_\tr{F}$ (left) and doping level $n_\tr{Gr}$ (right) in graphene as a function of the incident photon flux $\Phi_\tr{ph}$. Measurements on three selected Gr/MoSe$_2$ samples (denoted S$_1$, S$_2$ and S$_3$) are represented with different symbols. The data shown in Fig.\ref{Fig1}-\ref{FigMoSe2} is from S$_1$. The hatched region denotes the range of $E^\tr{Gr}_\tr{F}$ close to the charge neutrality point where there is an uncertainty on the sign and on the exact value of $E^\tr{Gr}_\tr{F}$. The gray rectangle illustrates the saturation value of $E^\tr{Gr}_\tr{F}$. (b) $E^\tr{Gr}_\tr{F}$ (left) and $n_\tr{Gr}$ (right) as a function of $\Phi_\tr{ph}$ obtained under ambient conditions (filled symbols) and in vacuum (open symbols) at the same point of a Gr/MoSe$_2$/SiO$_2$ and a MoSe$_2$/Gr/SiO$_2$ heterostructure, denoted S$_4$ and S$_5$, respectively. Measurements under vacuum in S$_4$ are shown as $\Phi_\tr{ph}$ is swept forward for the first time and then backward (see arrows). }
\label{FigAirVac}
\end{center}
\end{figure}

The charge density and exciton dynamics in 2D materials are known to be influenced by environmental effects, in particular  by molecular adsorbates and the underlying substrate~\cite{Ryu2010,Miller2015,Tongay2013,Cadiz2016}. To determine the generality of the results presented above, we compare in Fig.~\ref{FigAirVac}(a,b) the evolution of $E^\tr{Gr}_\tr{F}$ with increasing $\Phi_\tr{ph}$ recorded in ambient air and under high vacuum ($\lesssim 10^{-4}~\tr{mbar}$) for a set of five samples, wherein strong PL quenching has been observed (see Supplemental Material~\cite{SMnote}, Fig.~S11). Remarkably, in ambient air, all samples display (i) different initial doping at low $\Phi_\tr{ph}$, (ii)  distinct sub-linear rises of $E^\tr{Gr}_\tr{F}$ with increasing $\Phi_\tr{ph}$ and (iii) similar saturation at $E^\tr{Gr}_\tr{F}$ around $290\pm 15~\tr{meV}$ (i.e., $n_\tr{Gr}\approx (5\pm0.5)\times10^{12}~\tr{cm}^{-2}$). Interestingly, under vacuum, we systematically observe a transient regime with a photoinduced rise of $E^\tr{Gr}_\tr{F}$ (at fixed $\Phi_{\tr{ph}}$)  towards a saturation value that is attained on a rather long timescale (typically several minutes, depending on $\Phi_{\tr{ph}}$, see Fig.~\ref{FigAirVac}(b) and Supplemental Material~\cite{SMnote}, Fig.~S10). Once $E^\tr{Gr}_\tr{F}$ has reached its saturation value, it becomes completely independent on $\Phi_{\tr{ph}}$ (see Fig.~\ref{FigAirVac}(b)).


The distinct charge transfer dynamics observed under ambient conditions and in vacuum shed light on the role of molecular adsorbates at the surface of the vdWH. In vacuum, a significant fraction of the molecular adsorbates are removed, including through laser-assisted desorption. These adsorbates are efficient electron traps~\cite{Ryu2010,Miller2015,Tongay2013}, acting against the net photoinduced electron transfer from MoSe$_2$ to graphene. In the absence of molecular adsorbates, the electrons that are transferred from MoSe$_2$ to graphene remain on graphene as long as the laser illumination is on (see Sec.~\ref{Discussion}). Such extrinsic effects impact the optoelectronic response of 2DM and vdWH - most often examined under ambient conditions - and therefore provide an impetus for further studies under controlled atmospheres~\cite{Ryu2010}, using different substrates, stacking sequences and encapsulating materials~\cite{Raja2017,Ajayi2017,Cadiz2017}. Along this line, we have studied (see Fig.\ref{FigAirVac}(b) and Supplemental Material~\cite{SMnote}, Fig.~S11) a MoSe$_2$/Gr/SiO$_2$ vdWH. Remarkably, the results obtained on this \textit{inverted} heterostructure are very similar to those obtained in Gr/MoSe$_2$/SiO$_2$ vdWHs.

Finally, we have compared the PL in Gr/MoSe$_2$/SiO$_2$ and MoSe$_2$/Gr/SiO$_2$ in ambient air and under vacuum conditions. While the PL of bare MoSe$_2$ is -as previously reported~\cite{Tongay2013}- quenched under vacuum, the  PL intensity and lineshape measured as a function of $\Phi_{\tr{ph}}$ in ambient air and under vacuum in Gr/MoSe$_2$ are not appreciably different (See Fig.~\ref{FigICT-IET} and Supplemental Material~\cite{SMnote}, Fig.~S12).

\section{Discussion}
\label{Discussion}

The complementary results reported in Sec.~\ref{SecPL}-\ref{AirVac} now make it possible to address a set of open questions of high relevance for fundamental photophysics and optoelectronic applications. What are the microscopic mechanisms responsible for  net electron transfer and its saturation (Sec.~\ref{DICT})? What is the impact of excitonic effects on interlayer coupling (Sec.~\ref{DEXC})? What are the relative contributions of ICT and IET to the massive photoluminescence quenching analyzed in Fig.~\ref{FigPL} (Sec.~\ref{DICT-IET})?

\subsection{Charge transfer mechanism}
\label{DICT}

The clear saturation of the net photoinduced ICT in Gr/TMD heterostructures shown in Fig.~\ref{FigAirVac} had not been reported thus far and we shall first discuss the underlying microscopic ICT mechanisms. Since the Dirac point of graphene is located between the valence band maximum and the conduction band minimum of MoSe$_2$~\cite{Yu2009,Liang2013,Wilson2017}, the tunneling of photoexcited electrons \textit{and} holes to graphene can be envisioned as long as energy and momentum are conserved and that $E^\tr{Gr}_\tr{F}$ lies sufficiently below (above) the conduction band minimum (valence band maximum) of MoSe$_2$. Electron and hole transfer to graphene are sketched in Fig.~\ref{FigSketch}(a,b). To account for our experimental findings, we propose the following scenario.

The band structure of coupled Gr/MoSe$_2$ can be, in first approximation, considered as the superposition of the bands of the different materials~\cite{Kosmider2013,Pierucci2016b} separated by a subnanometer ``van der Waal gap''. The relative position of the band structure is determined by the offsets between the Dirac point of graphene and the valence (conduction) band maximum (minimum) of MoSe$_2$.  In the dark, without loss of generality we may assume that both graphene and MoSe$_2$ are quasi-neutral. When visible light is shined onto Gr/MoSe$_2$, electron-hole pairs and excitons  are mainly created in MoSe$_2$ since the latter absorbs significantly more than graphene~\cite{Mak2012,Li2015}. At this point, given the very close electron and hole effective masses in MoSe$_2$~\cite{Kormanyos2015}, the rates of photoinduced electron and hole transfer from MoSe$_2$ to graphene will chiefly depend on the the wavefunction overlap, the density of states in graphene and the energy difference between the band extrema in MoSe$_2$ and $E^\tr{Gr}_\tr{F}$.

Assuming the Dirac point lies closer to the valence band maximum than to the conduction band minimum~\cite{Yu2009,Liang2013},  the photoinduced electron current to graphene should exceed the hole current immediately after sample illumination, consistently with our experimental findings. Due to the  small density of states of graphene near its Dirac point~\cite{Castroneto2009}, the net electron transfer to graphene induces a sizeable rise of $E^\tr{Gr}_\tr{F}$ above the Dirac point. Thus, as $E^\tr{Gr}_\tr{F}$ moves away from the valence band maximum in MoSe$_2$, the hole current to $n$-doped graphene increases significantly. The vanishing of the net electron transfer to graphene then results from the cancellation of the photoinduced electron (Fig.~\ref{FigSketch}(a)) and hole (Fig.~\ref{FigSketch}(b)) currents. 
In vacuum and in the absence of adsorbates, this saturation is reached in the steady state at any $\Phi_\tr{ph}$. In ambient air, electrons may escape from graphene (in Gr/MoSe$_2$/SiO$_2$) or MoSe$_2$ (in MoSe$_2$/Gr/SiO$_2$) resulting (at intermediate $\Phi_\tr{ph}<10^{23}~\tr{cm}^{-2}~\tr{s}^{-1}$) in a steady state $E^\tr{Gr}_\tr{F}$ below the $\Phi_\tr{ph}$-independent saturation value observed in vacuum (see Fig.~\ref{FigAirVac}(b))~\footnote{As a result, the relative magnitudes of the electron and hole flows, and the resulting steady state $E^\tr{Gr}_\tr{F}$ are not exclusively determined by $\Phi_\tr{ph}$ (compare data in Fig.~\ref{FigAirVac} and see Supplemental Material~\cite{SMnote}, Fig. S13).}.

The very similar saturation values of $E^\tr{Gr}_\tr{F}$ uncovered in several Gr/MoSe$_2$ samples both in ambient air \textit{and} in vacuum (see Fig.~\ref{FigAirVac}) suggest an limit set by the intrinsic band offsets between graphene and MoSe$_2$, as well as the electron and hole tunnelling efficiencies. The latter be affected by extrinsic materials properties, such as the presence of band tails states as well as other traps and defects~\cite{Furchi2014}. Systematic studies using other TMDs with distinct band offsets relative to graphene, and controlled amounts of impurities and/or defects will help determining the shares of extrinsic and intrinsic effects in the net charge transfer saturation. Nevertheless, our work is a step towards optical determination of band offsets in van der Waals heterostructures. Confronted to electron transport measurements~\cite{Kim2015c}, or angle-resolved photoemission spectroscopy~\cite{Pierucci2016b,Wilson2017}, our Raman-based approach may unveil the impact of strong bandgap renormalization and exciton binding energy on the optoelectronic properties of TMDs and related vdWHs.

 \subsection{Impact of excitonic effects}

\begin{figure}[!t]
\begin{center}
\includegraphics[width=0.9\linewidth]{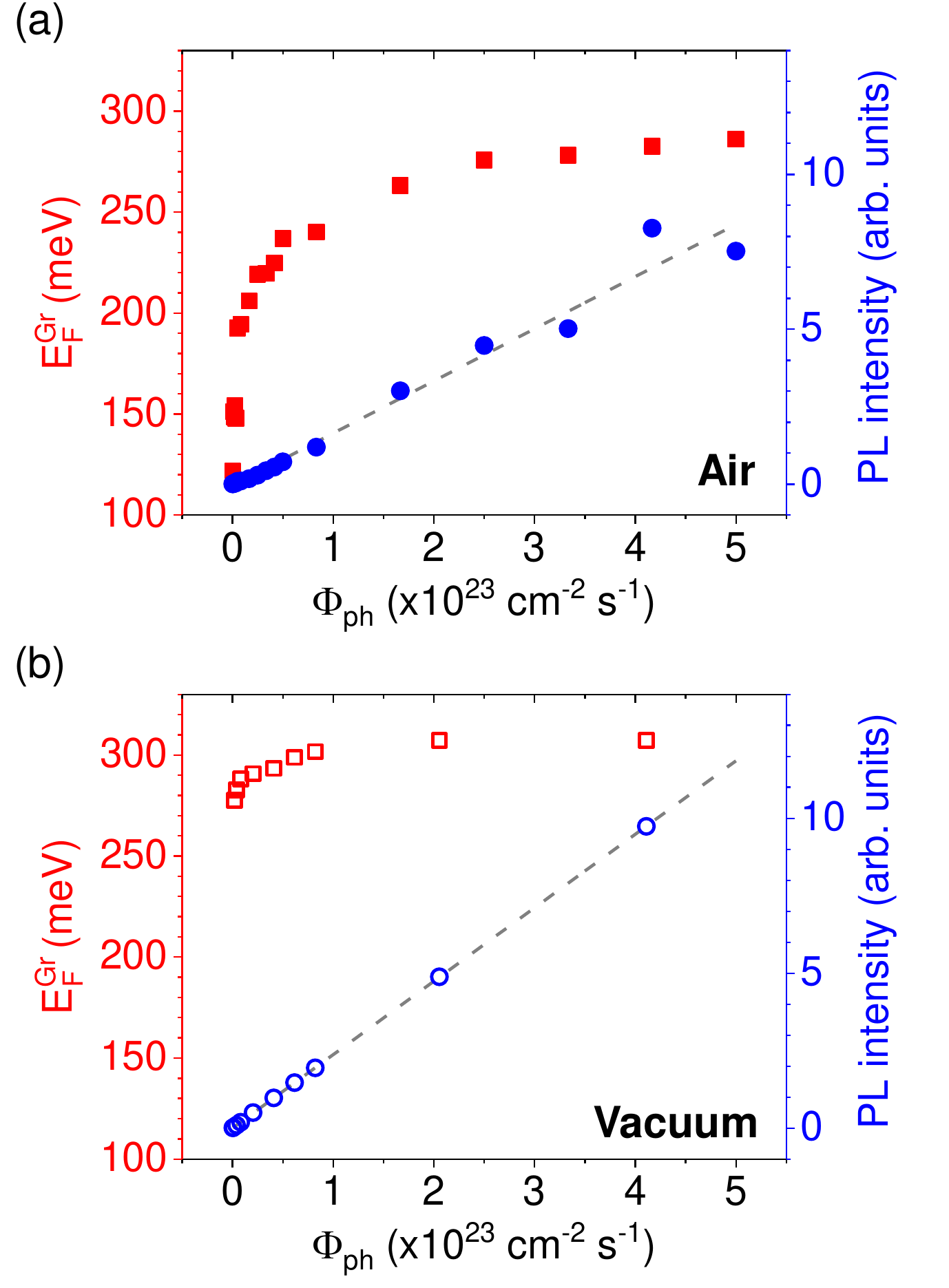}
\caption{ Comparison between the dependence of $E^\tr{Gr}_\tr{F}$ (left axis, red) and $I_\tr{PL}^{\tr A}$ (right axis, blue) on $\Phi_\tr{ph}$ measured on sample S$_2$ (a) in ambient air and (b) in vacuum at a laser photon energy of 2.33~eV. The dashed lines are linear fits to the photoluminescence data.}
\label{FigICT-IET}
\end{center}
\end{figure}

\begin{figure*}[!t]
\begin{center}
\includegraphics[width=0.85\linewidth]{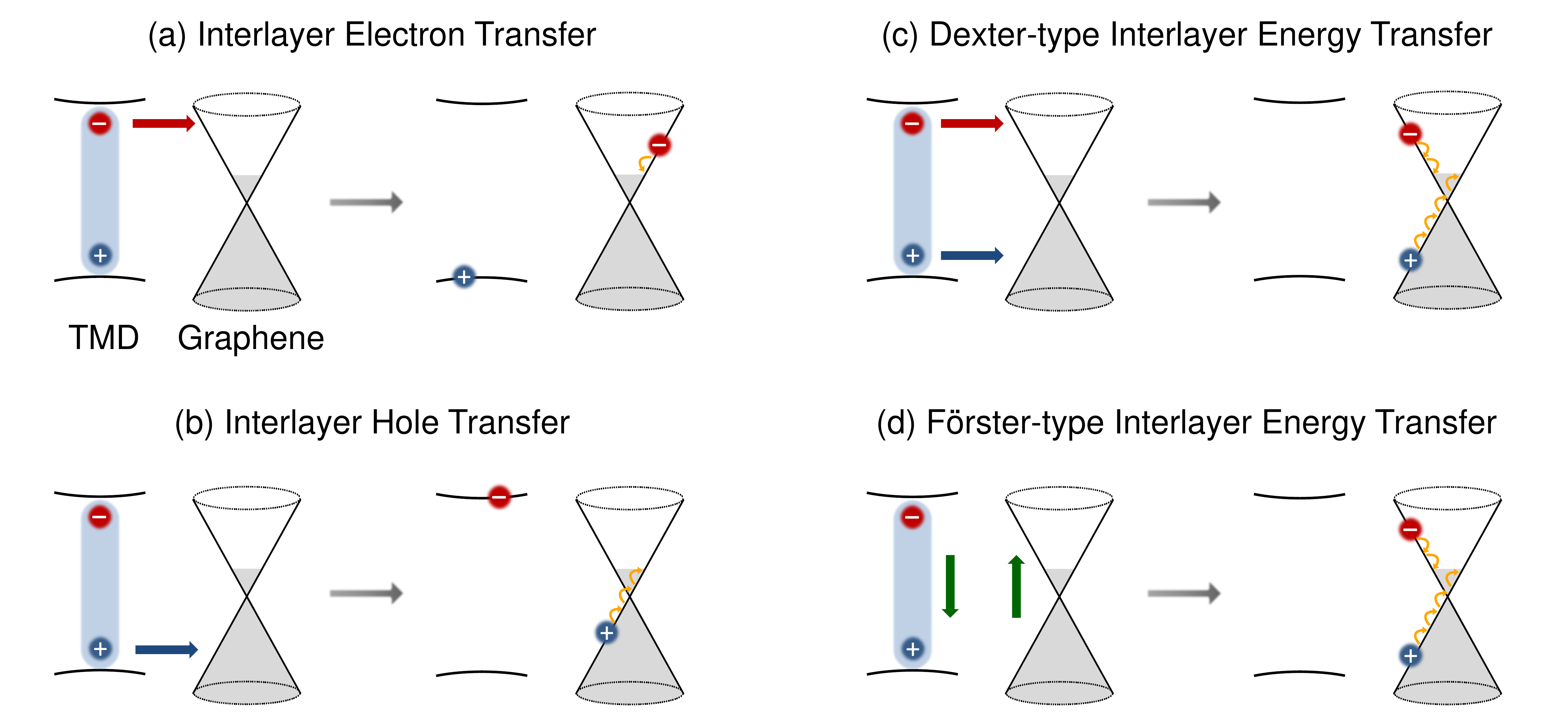}
\caption{Sketches, in momentum-energy space, of interlayer (a) electron and (b) hole transfer, (c) Dexter-type and (d) F\"orster-type energy transfer processes from monolayer TMD to graphene. Band-edge (A) excitons are symbolized by the blue shaded area with electron in red and hole in dark blue. The curved dark yellow arrows represent ultrafast energy relaxation of transferred carriers in graphene down to the Fermi level. In the case of balanced electron and hole flows to graphene, the charge transfer processes effectively result in energy transfer.}
\label{FigSketch}
\end{center}
\end{figure*}

\label{DEXC}
 Indeed, although PL measurements (see Fig.~\ref{FigPL}) make it clear that excitons are formed in MoSe$_2$, the impact of excitonic effects on interlayer coupling and more generally on the optoelectronic response of vdWH remains elusive.

Upon optical excitation well-beyond the optical bandgap (as it is the case in Fig.~\ref{Fig1}-\ref{FigICT-IET}), free electron-hole pairs \textit{and} tightly bound excitons can be formed in Gr/MoSe$_2$~\cite{Raja2017,Ugeda2014}.  Despite exciton formation being highly efficient and occurring on sub-picosecond to a few picosecond timescales~\cite{Robert2016,Steinleitner2017}, our PL measurements have revealed  equally short band-edge (A) exciton lifetimes in Gr/MoSe$_2$ (see Fig~\ref{FigPL}). Therefore the observed interlayer coupling processes may certainly involve band-edge excitons but may also imply direct \textit{hot} carrier and/or higher-order exciton transfer to graphene. To assess the contribution of out of equilibrium effects, we have combined PL and Raman measurements in ambient air and in vacuum on a Gr/WS$_2$/SiO$_2$ vdWH at two different laser photon energies near the B exciton (2.33~eV) and slightly below the A exciton (1.96~eV). In the latter case only A excitons can be formed by means of an upconversion process~\cite{Jones2016,Chervy2017} (see Supplemental Material~\cite{SMnote}, Fig.~S14). For both incoming photon energies, we observe strong PL quenching as well as photoinduced doping, very similar to the observations discussed in Fig.~\ref{FigPL}-\ref{FigICT-IET} for Gr/MoSe$_2$ vdWHs. These observations indicate that ICT and IET  processes in GR/TMD vdWH mainly involve band-edge TMD excitons \footnote{This conclusion is consistent with the fact that momentum conservation can be more easily fulfilled for an exciton than for a free charge carrier. Indeed, an exciton in the TMD can decay by transferring an electron or hole to a finite momentum state graphene leaving the other carrier in the TMD with the excess momentum (see Fig.~\ref{FigSketch}(a,b)), whereas at room temperature, a free charge carrier near the band-edges would need extra momentum provided by defect or phonon scattering.}. This result illustrates the unusually strong excitonic effects in TMDs, which must be taken into consideration when adapting free-carrier optoelectronic models to the case of vdWH-based devices.

 \subsection{Charge vs energy transfer}
\label{DICT-IET}

Finally, we address the competition between ICT and IET. Let us first recall that Raman measurements probe the steady state charge carrier densities in our samples and do not make it possible to extract electron and hole transfer rates. Figure~\ref{FigICT-IET} summarizes our findings by confronting the dependence of $E^\tr{Gr}_\tr{F}$ and $I^\tr{A}_\tr{PL}$ on $\Phi_{\tr{ph}}$ in sample S$_2$. The key implications of our combined PL and Raman study are that the short exciton lifetime in Gr/MoSe$_2$ is (i) independent on $\Phi_{\tr{ph}}$ (over nearly four orders of magnitude), (ii) unaffected by  the environmental conditions (air vs vacuum), and, crucially by (iii) the presence (in air, at low $\Phi_{\tr{ph}}$) or absence (in vacuum at any $\Phi_{\tr{ph}}$, or in air at high $\Phi_{\tr{ph}}$) of \textit{net} photoinduced ICT (here, electron transfer from MoSe$_2$ to graphene, see also Supplemental Material~\cite{SMnote}, Fig.~S13).

Our data demonstrate that albeit electrons and holes may transfer to graphene, ICT processes alone (even in the case of balanced electron and hole transfer) cannot be responsible for the massive PL quenching and linear rise of the PL intensity vs $\Phi_{\tr{ph}}$. Instead, IET -- either in the form of electron exchange (i.e., Dexter-type IET, Fig.~\ref{FigSketch}(c)) or dipole-dipole interaction (i.e., F\"orster-type IET, Fig.~\ref{FigSketch}(d) -- provides a highly efficient relaxation pathway for excitons in Gr/TMD heterostructures.

Consequently, the $\sim 1~\rm ps$ exciton lifetime deduced from PL measurements (see Sec.~\ref{SecPL}) can be considered as a fair estimation of the energy transfer timescale from a TMD monolayer to a graphene monolayer placed in its immediate vicinity. Since interlayer coupling is highly sensitive to minute changes (at the \angstrom~level) in the distance between 2D layers as well as to the distribution of TMD excitons in energy-momentum space~\cite{Federspiel2015,Robert2016}, a timely theoretical and experimental challenge is to unravel the distance and temperature dependence of the energy transfer rate and quantify contributions stemming from short range (Dexter) and longer range Förster mechanisms.

Let us add the following comments in order to tentatively pinpoint the microscopic energy transfer mechanism. First, although balanced ICT and Dexter-type IET follow \textit{a priori} two distinct microscopic mechanisms (see Fig.~\ref{FigSketch}(a-c)), both processes imply charge tunnelling (i.e., wavefunction overlap) and result in a similar final state where the energy of an exciton population is transferred to graphene. Interestingly, it was recently predicted in porphyrin/graphene hybrids that Dexter ET is largely inefficient compared to F\"orster ET even at sub-nanometer distances~\cite{Malic2014}. In the case of Gr/TMD vdWHs, the large in-plane dipoles in monolayer TMDs~\cite{Schuller2013} should further favor F\"orster energy transfer to graphene. Along this line, the exciton lifetime measured in \textit{decoupled} Gr/MoSe$_2$/SiO$_2$ (see Fig.~\ref{FigPL}(e)) is of the same order of magnitude yet appreciably shorter than in MoSe$_2$/SiO$_2$, an effect that may tentatively be assigned to long-range F\"orster energy transfer~\footnote{Let us note that the PL data in Fig.~\ref{Fig1}(f) and Fig.~\ref{FigPL}(e) have been recorded on freshly made sample S1, before the data in Fig.~\ref{FigPL}(c)-(d). Aging of the air-exposed MoSe$_2$ layer in S1 is likely responsible for the fact that in Fig.~\ref{FigPL}(c)-(d), the PL intensity in MoSe$_2$/SiO$_2$ is slightly smaller than in decoupled Gr/MoSe$_2$/SiO$_2$, wherein graphene acts as an efficient passivating layer.}.

\section{Conclusion and outlook} 


We have exploited complementary insights from micro-Raman and photoluminescence spectroscopies to disentangle contributions from interlayer charge and energy transfer in graphene/TMD heterostructures and establish the key role of energy transfer. These general findings advance our fundamental understanding of light-matter interactions at atomically-thin heterointerfaces and have far reaching consequences for applications.

Indeed the Gr/TMD system is a ubiquitous building block in emerging optoelectronic nanodevices. Having established that edge TMD excitons transfer to graphene with near-unity efficiency, a key challenge is now to separate the electron-hole pairs formed in graphene~\cite{Brenneis2015} and enhance photoconductivity and/or photocurrent generation before these charge carriers release their energy into heat on a sub-picosecond timescale\cite{Johannsen2013,Gierz2013}.

The competition between interlayer charge and energy transfer is also a matter of active debate in related systems, e.g. in TMD/TMD type II heterojunctions~\cite{Hong2014,Ceballos2014,Fang2014,Lee2014a,Rivera2015,Kozawa2016}, that are also of high relevance for optoelectronics~\cite{Mak2016} and valleytronics~\cite{Schaibley2016}. We have shown that fingerprints of interlayer charge transfer are encoded in the Raman response of TMD monolayers. Combining Raman measurements and photoluminescence spectroscopy of intra- and inter-layer excitons should provide decisive insights into exciton dynamics in these atomically-thin semiconductor heterostructures.

More broadly, van der Waals heterostructures are also emerging as a platform to explore many-body effects and new regimes of strong- and/or chiral light-matter interactions. Further developments in these emerging areas will benefit from the present insights into interlayer charge and energy transfer.

\appendix
\section{Experimental details} Gr/MoSe$_2$ vdWHs were prepared onto Si wafers covered with a 90 nm thick SiO$_2$ epilayer using a viscoelactic transfer technique~\cite{Castellanos2014}. The Gr/MoSe$_2$ vdWHs were first characterized by atomic force microscopy (AFM) and then by micro-PL and micro-Raman measurements. As-prepared samples (such as sample $S_1$ discussed above, see Fig.~\ref{Fig1}(a)) as well as annealed samples (1 hour at 150$^{\circ}\tr{C}$ and 2 hours at 200$^{\circ}\tr{C}$ in high vacuum) such as sample S$_3$ were studied. Although more ``pockets'' (see Supplemental Material~\cite{SMnote}, Fig.~S1) are present on as-prepared samples, we could observe, both in annealed and as-prepared samples, extended ($>25~\mu \tr m^2$) \textit{coupled} Gr/MoSe$_2$ domains with smooth and uniform interfaces due to a self-cleaning process~\cite{Haigh2012}.

PL and Raman studies were carried out in ambient air and in high vacuum, in a backscattering geometry, using a home-built confocal microscope. Unless otherwise noted, the samples were optically excited using a single longitudinal mode, linearly polarized, $2.33~\tr{eV}$ ($532~\tr{nm}$) continuous wave laser. The collected light was dispersed onto a charged-coupled device (CCD) array by a single (500~nm in focal length) monochromator equipped with a 150 (resp. 900 for graphene, 2400 for MoSe$_2$) grooves/mm grating for PL (resp. Raman) measurements. Spectral resolutions of $1.4~\tr{cm}^{-1}$ and $0.6~\tr{cm}^{-1}$ were obtained for Raman measurements with the 900 and 2400 grooves/mm grating, respectively. The sample holder was mounted onto a x-y-z piezoelectric stage, allowing hyperspectral imaging. Time-resolved PL measurements were performed on the same setup using a pulsed supercontinuum laser, with a repetition rate tunable from $1.95~\tr{MHz}$ up to $78~\tr{MHz}$. The unpolarized output of the supercontinuum laser at $1.96~\tr{eV}$ ($633~\tr{nm}$) was selected using an acousto-optic tunable filter. PL decays were obtained using an avalanche photodiode coupled to a time-tagged, time-correlated single photon counting board. We have employed a very broad range of photon fluxes resulting in exciton densities below and beyond the values achieved in earlier works \cite{He2014,Massicotte2016,Zhang2014}. The incident photon flux $\Phi_\tr{ph}$ is obtained by measuring the laser power and the area of the laser spot. For instance, with a measured optical power of $1~\mu\tr{W}$ at the objective at 2.33~eV, we obtain a photon flux of $2.2\times10^{21}~\tr{cm}^{-2}~\tr{s}^{-1}$ using a 100x objective with a numerical aperture of 0.9. 
The Raman peaks are fit using Lorentzian and modified Lorentzian~\cite{Froehlicher2015a,Metten2014} profiles for the G- and 2D-mode features, respectively, and Voigt profiles (with a fixed Gaussian width taking into account our spectral resolution) for MoSe$_2$ features. Therefore in Fig.~\ref{FigMoSe2}(b), the linewidth $\Gamma_{A^{\prime}_1}$ has to be understood as the Lorentian FWHM extracted from a Voigt fit.

\begin{acknowledgments}
 We thank D.M. Basko, C. Robert, D. Lagarde, X. Marie, A. Ouerghi, G. Schull and C. Genet for fruitful discussions. We are grateful to J. Chr\'etien for his early contribution to experimental measurements, to H. Majjad and M. Rastei for help with AFM measurements, and to the StNano clean room staff and M. Romeo for technical assistance. We acknowledge financial support from the Agence Nationale de la Recherche (under grant H2DH ANR-15-CE24-0016) and from the LabEx NIE (under Grant ANR-11-LABX-0058-NIE within the Investissement d’Avenir program ANR-10-IDEX-0002-02). S.B. is a member of Institut Universitaire de France (IUF). 
\end{acknowledgments}




%


\onecolumngrid
\newpage
\begin{center}
{\LARGE\textbf{Supplemental Material}}
\end{center}

\setcounter{equation}{0}
\setcounter{figure}{0}
\setcounter{section}{0}
\renewcommand{\theequation}{S\arabic{equation}}
\renewcommand{\thefigure}{S\arabic{figure}}
\renewcommand{\thesection}{S\arabic{section}}
\renewcommand{\thesubsection}{S\arabic{section}.\arabic{subsection}}
\renewcommand{\thesubsubsection}{S\arabic{section}.\arabic{subsection}.\arabic{subsubsection}}
\linespread{1.4}

\bigskip

\section{Additional Results on Sample S$_1$}

\subsection{Atomic force microscopy}

\begin{figure}[!tbh]
\begin{center}
\includegraphics[width=0.9\linewidth]{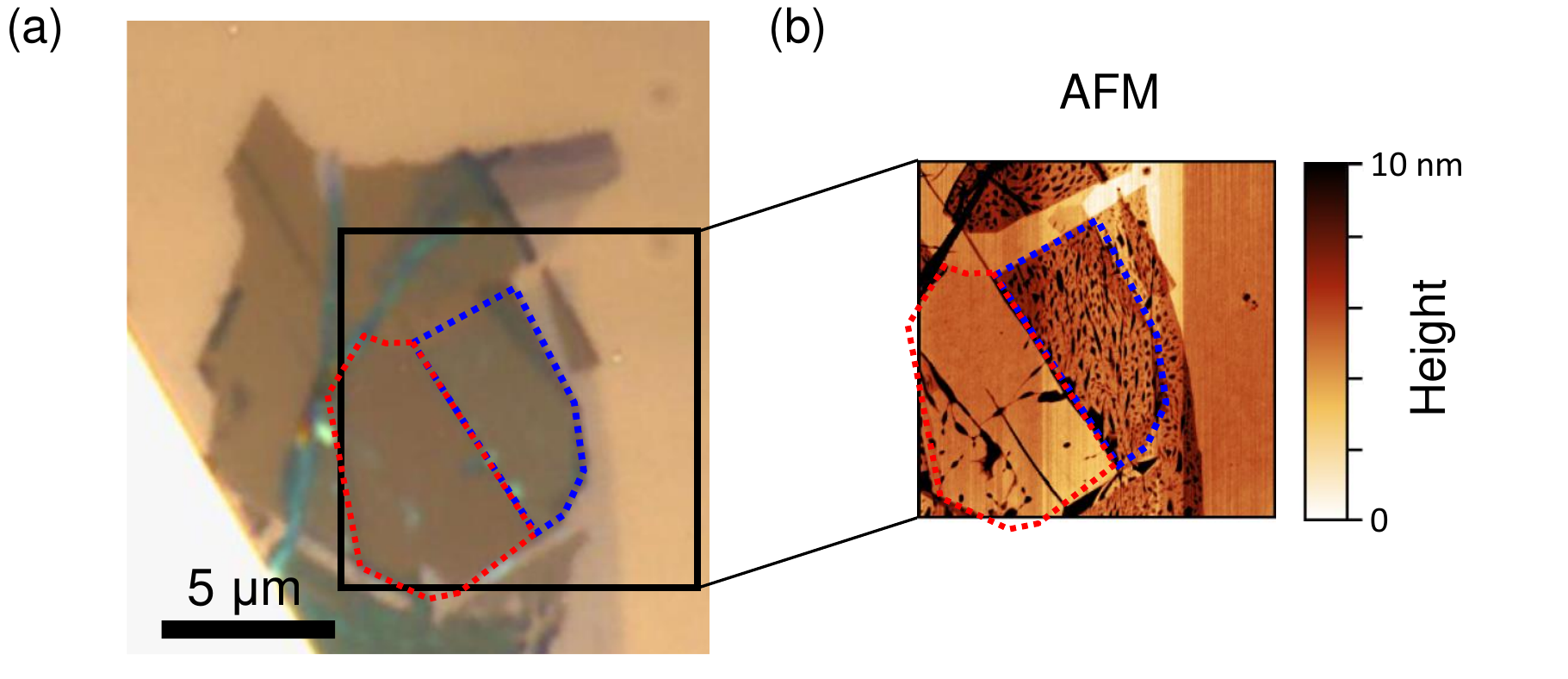}
\caption{(a) Optical image of the sample presented in the main text (denoted S1 in Fig. 4). (b) Atomic force microscopy image of the black square in (a). The coupled and decoupled regions are highlighted with red and blue dashed lines, respectively.  In the coupled part, the interface between the two layers is free of contamination and is atomically flat due to the so-called ``self-cleaning'' mechanism~\cite{Haigh2012,Kretinin2014}. On the other hand, the interface in the decoupled part shows contamination pockets.}
\label{FigSI1}
\end{center}
\end{figure}

\clearpage

\subsection{Graphene Raman spectra for increasing $\Phi_\tr{ph}$}

\begin{figure}[!tbh]
\begin{center}
\includegraphics[width=1\linewidth]{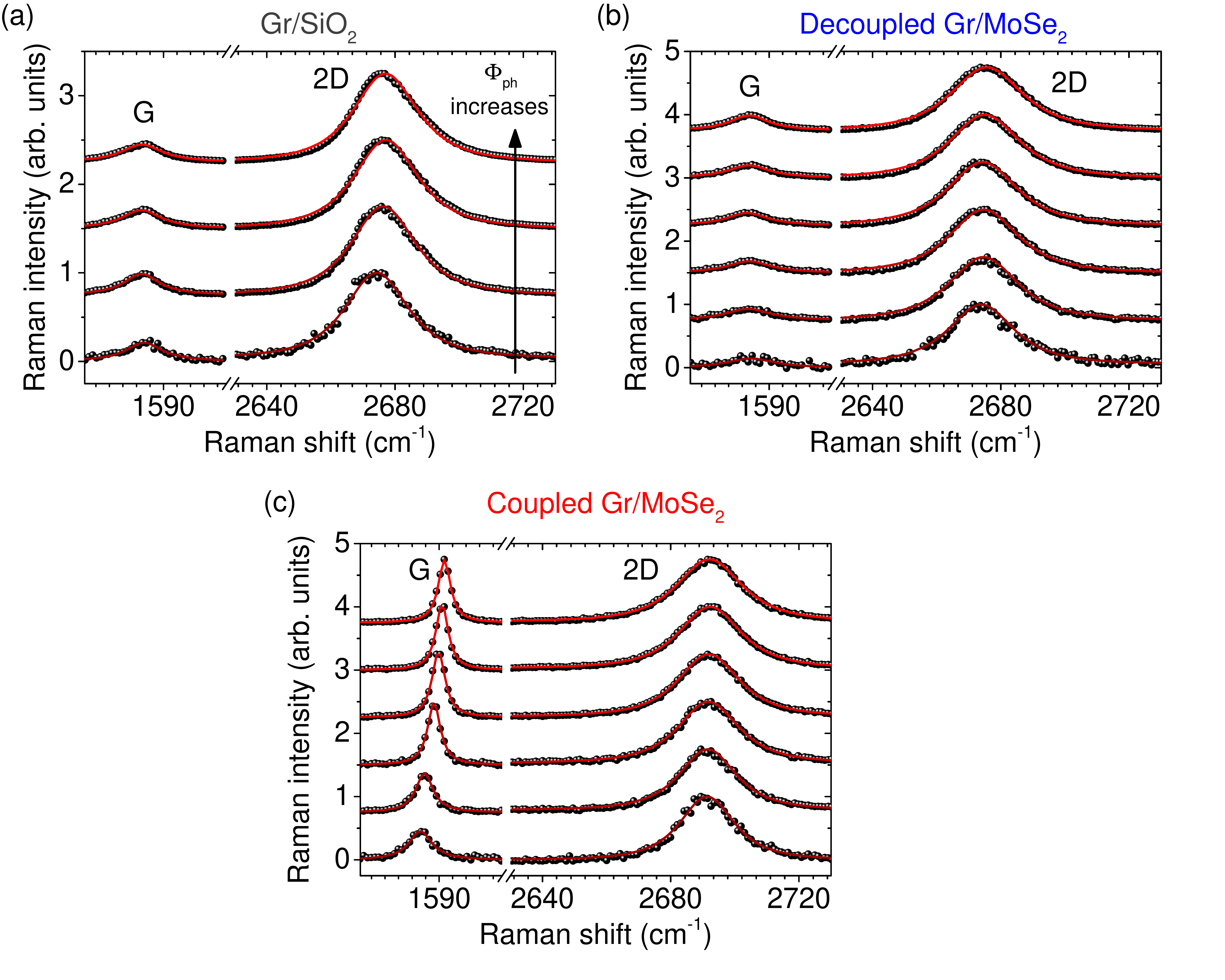}
\caption{Raman spectra corresponding to the data in Fig. 2. Measurements are performed  in ambient conditions, at a photon energy $E_\tr{L}=2.33~\tr{eV}$, for increasing values of incident photon flux ($\Phi_\tr{ph}$), between $3.3\times 10^{20}~\tr{cm}^{-2}~\tr{s}^{-1}$ and $4.2\times 10^{23}~\tr{cm}^{-2}~\tr{s}^{-1}$, for (a) Gr/SiO$_2$, (b) decoupled and (c) coupled Gr/MoSe$_2$. The spectra are vertically offset for clarity. Symbols are the experimental data and the solid lines are Lorentzian (G mode) and modified Lorentzian~\cite{Basko2008,Berciaud2013} (2D mode) fits. A broad Lorentzian background has been subtracted from the G-mode spectra. We observe that the Raman spectra of Gr/SiO$_2$ and decoupled Gr/MoSe$_2$ are not affected by the increase of $\Phi_\tr{ph}$, whereas the Raman spectra of coupled Gr/MoSe$_2$ reveal clear fingerprints of photoinduced electron transfer (see main text).}
\label{FigSI2}
\end{center}
\end{figure}

\clearpage

\subsection{MoSe$_2$ Raman spectra for increasing $\Phi_\tr{ph}$}

\subsubsection{$A^{\prime}_1$ mode}

\begin{figure}[!tbh]
\begin{center}
\includegraphics[width=0.95\linewidth]{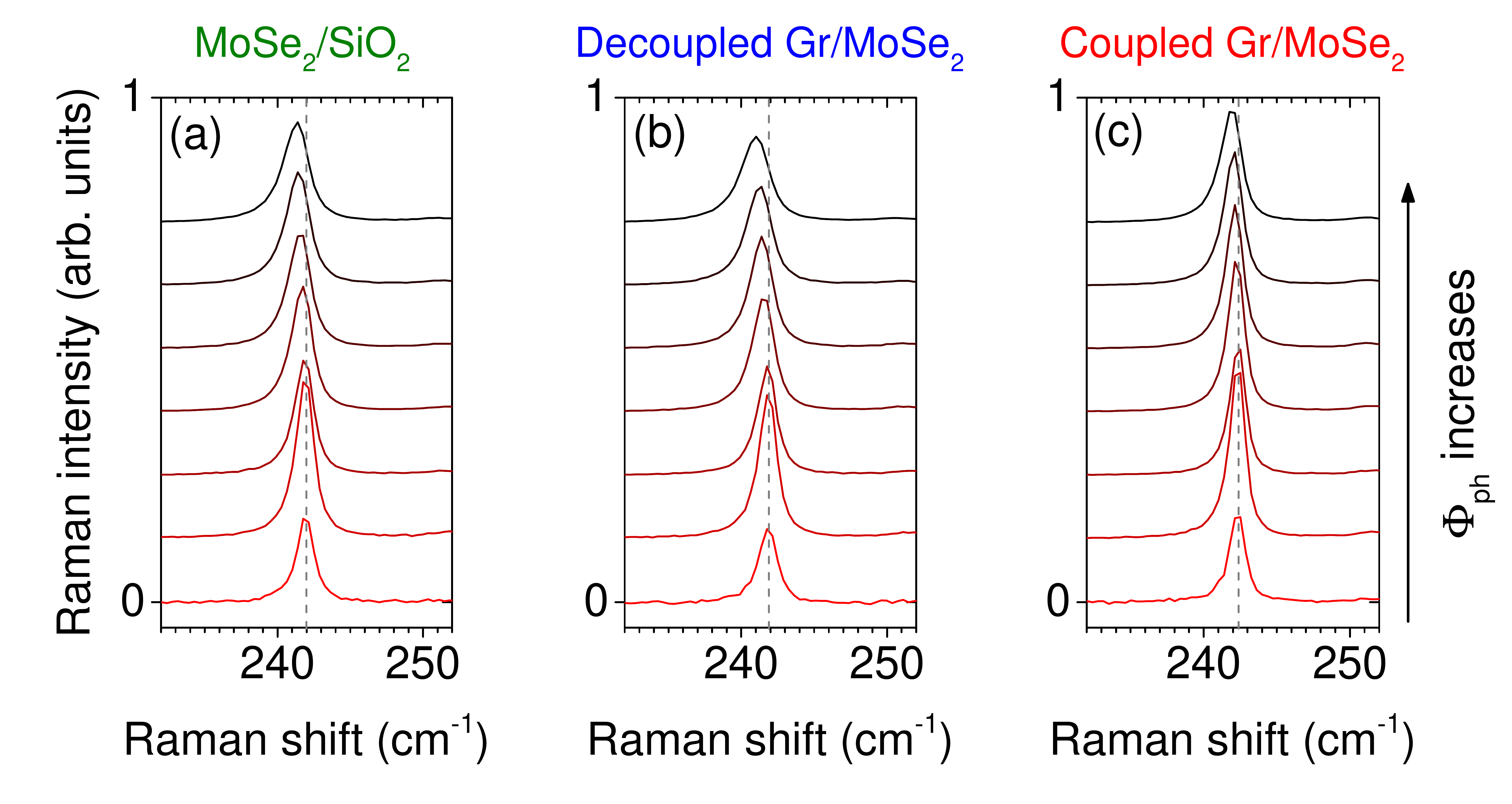}
\caption{Raman spectra of the $A^{\prime}_1$-mode feature in MoSe$_2$ recorded  in ambient conditions at $E_\tr{L}=2.33~\tr{eV}$ for increasing values of incident photon flux ($\Phi_\tr{ph}$), between $3.3\times 10^{20}~\tr{cm}^{-2}~\tr{s}^{-1}$ and $6.7\times 10^{23}~\tr{cm}^{-2}~\tr{s}^{-1}$ for (a) Gr/SiO$_2$, (b) decoupled and (c) coupled Gr/MoSe$_2$. The spectra are vertically offset for clarity. The vertical gray dashed lines indicate the frequency measured at the lowest $\Phi_\tr{ph}$.}
\label{FigSI3}
\end{center}
\end{figure}

\clearpage

\subsubsection{$E^{\prime}$ mode}

\begin{figure}[!tbh]
\begin{center}
\includegraphics[width=0.85\linewidth]{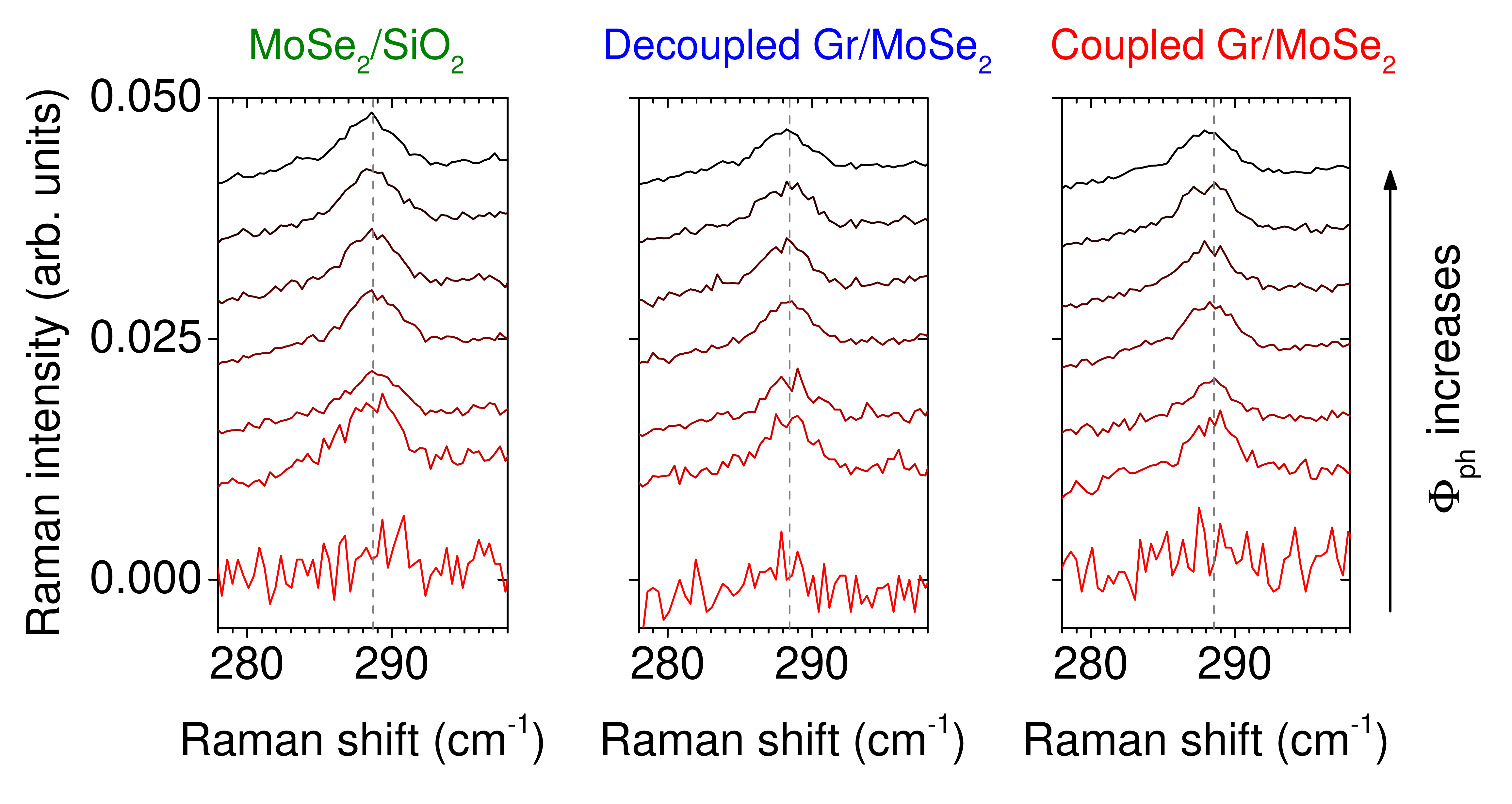}
\caption{Raman spectra of the $E^{\prime}$-mode feature in MoSe$_2$ recorded  in ambient conditions at $E_\tr{L}=2.33~\tr{eV}$ for increasing values of incident photon flux ($\Phi_\tr{ph}$), between $3.3\times 10^{20}~\tr{cm}^{-2}~\tr{s}^{-1}$ and $6.7\times 10^{23}~\tr{cm}^{-2}~\tr{s}^{-1}$ for (a) Gr/SiO$_2$, (b) decoupled and (c) coupled Gr/MoSe$_2$. The spectra are vertically offset for clarity. The vertical gray dashed lines indicate the frequency measured at the second lowest $\Phi_\tr{ph}$.}
\label{FigSI4}
\end{center}
\end{figure}

\begin{figure}[!tbh]
\begin{center}
\includegraphics[width=0.9\linewidth]{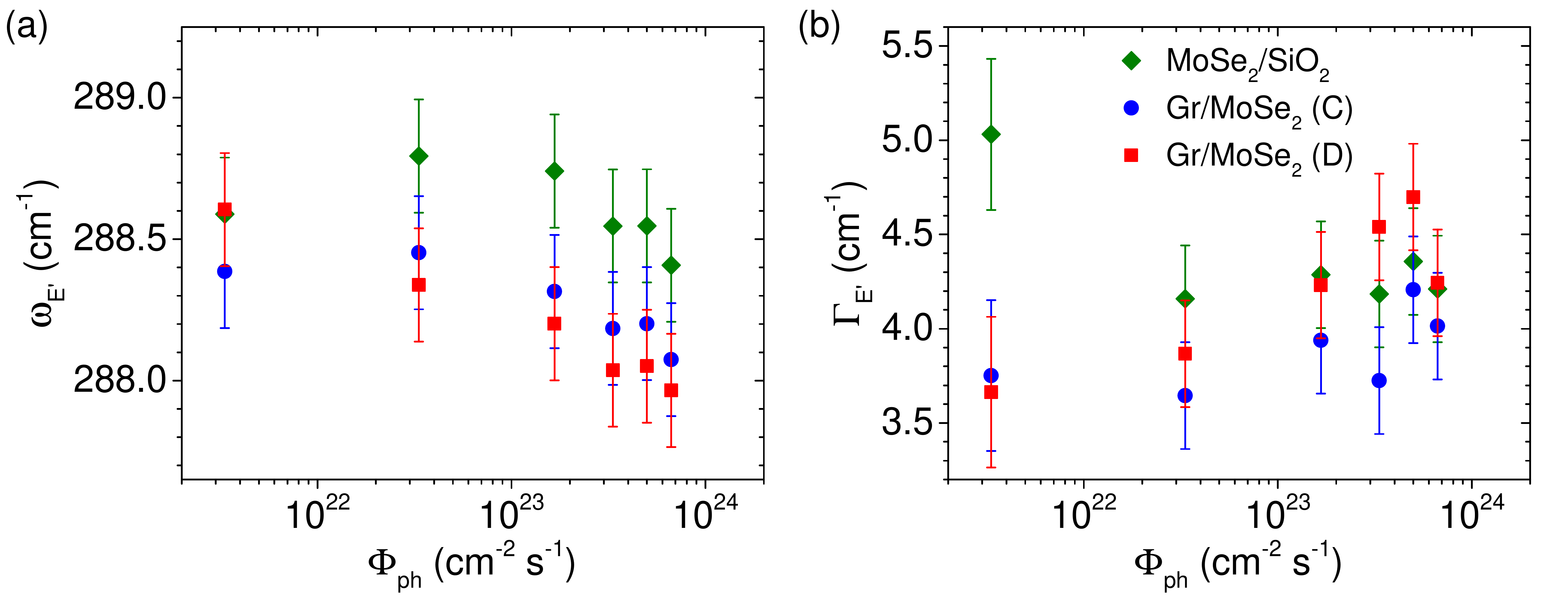}
\caption{Extracted (a) frequency $\omega_{E^{\prime}}$ and (b) FWHM $\Gamma_{E^{\prime}}$ from Fig.~\ref{FigSI8} as a function of the incident photon flux $\Phi_\tr{ph}$. Note that the spectra at $3.3\times 10^{20}~\tr{cm}^{-2}~\tr{s}^{-1}$ were not fit due to a weak signal. No significant changes with $\Phi_\tr{ph}$ are observed.}
\label{FigSI5}
\end{center}
\end{figure}

\clearpage

\subsection{Spatially-resolved Raman studies}

\subsubsection{Two-dimensional maps}

\begin{figure}[!h]
\begin{center}
\includegraphics[width=0.72\linewidth]{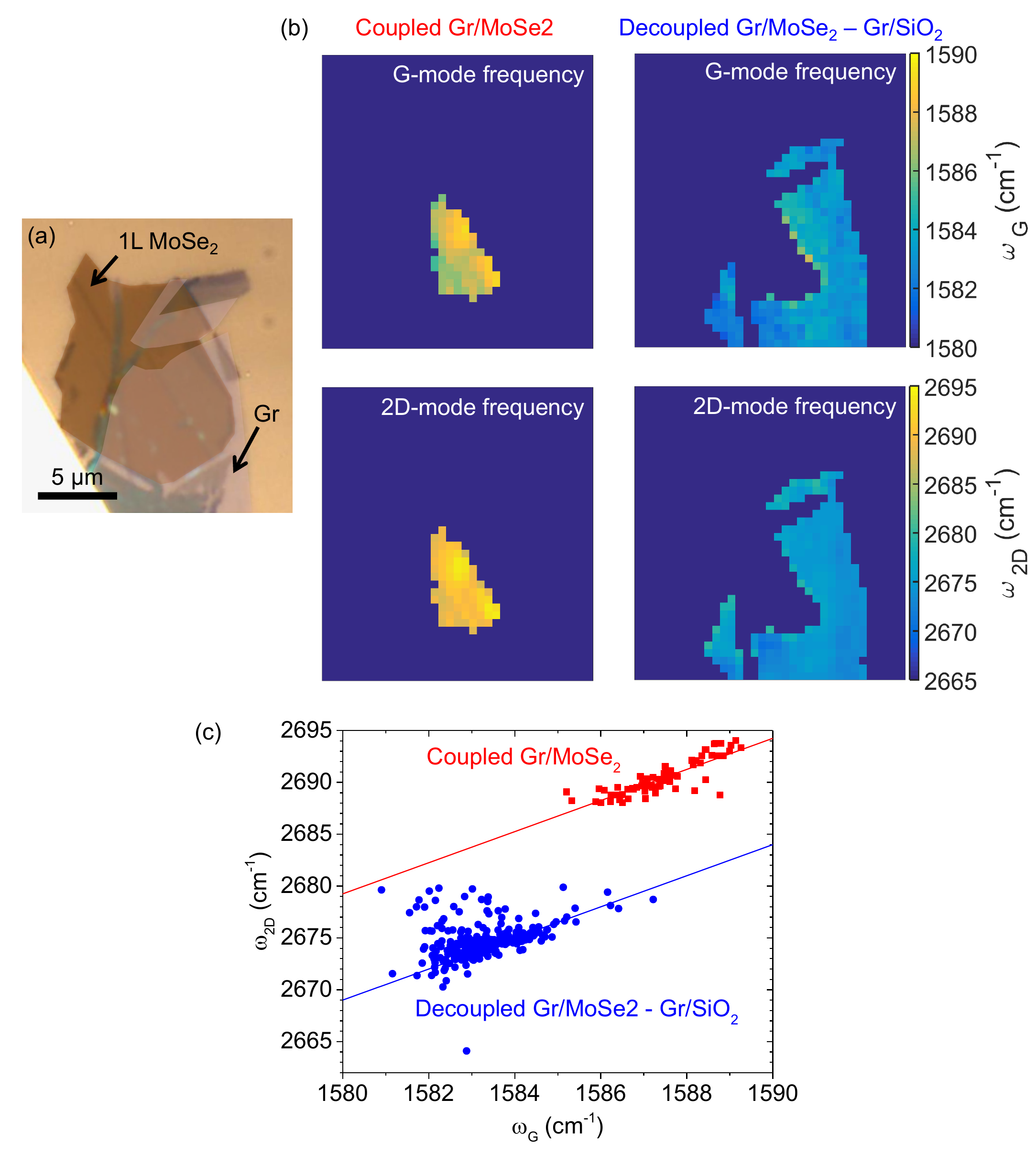}
\caption{(a) Optical image of sample 1. (b) G-mode frequency $\omega_\tr{G}$ and 2D-mode frequency $\omega_\tr{2D}$ for the coupled part of the heterostructure (left) or for the rest of the graphene monolayer (either deposited on SiO$_2$ or decoupled from MoSe$_2$, right). (c) Correlations between the frequencies of the 2D- and G-mode features shown in (b). The $(\omega_{\rm 2D},\omega_{\rm G})$ points cluster around mean values of $(2674.7\pm1.8~\rm cm^{-1},1583.3\pm0.9~\rm cm^{-1})$ on decoupled Gr/MoSe$_2$ and Gr/SiO$_2$ (blue circles), and $(2690.4\pm1.7~\rm cm^{-1}$,$1587.4\pm0.9~\rm cm^{-1})$ on coupled Gr/MoSe$_2$ (red squares). Their dispersions around these mean values follow linear correlations with a same slope of $\approx 1.5$, that suggests the coexistence of both a native strain field (leading to a slope of $\approx 2.2$)~\cite{Lee2012,Metten2014} and unintentional doping heterogeneities (leading to a slope of $\approx 0.1-0.6$)~\cite{Lee2012,Froehlicher2015a}.}
\label{FigSI6}
\end{center}
\end{figure}

\clearpage

\subsubsection{Line scans at various incident photon fluxes}

\begin{figure}[!tbh]
\begin{center}
\includegraphics[width=0.85\linewidth]{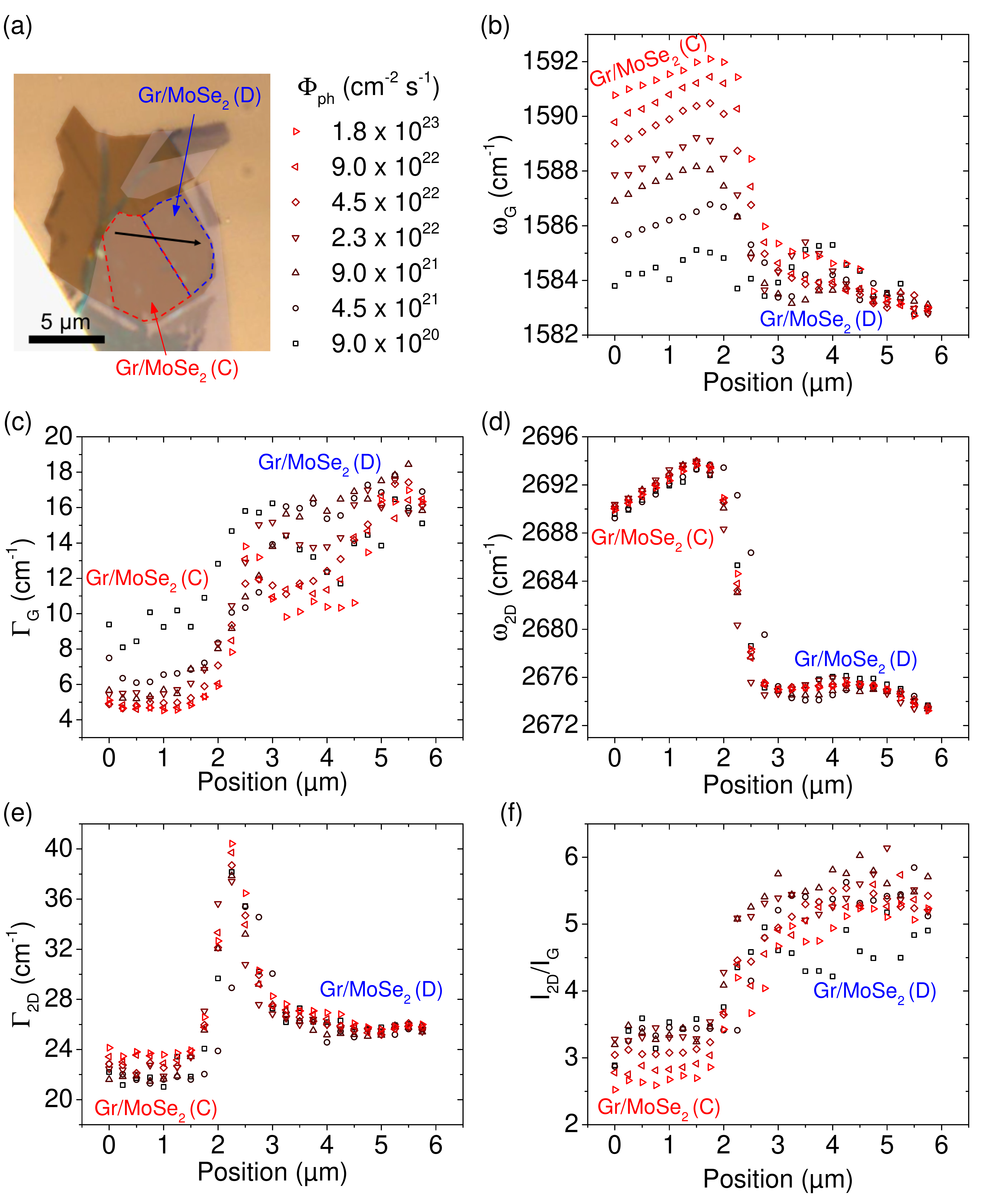}
\caption{(a) Optical image of sample S1. The red and blue dashed contours delimits the coupled and decoupled Gr/MoSe$_2$ regions. The line scans were recorded along the black arrow. The incident photon flux $\Phi_\tr{ph}$ are indicated. (b) G-mode frequency $\omega_\tr{G}$, (c) G-mode FWHM $\Gamma_\tr{G}$, (d) 2D-mode frequency $\omega_\tr{2D}$, (e) 2D-mode FWHM $\Gamma_\tr{2D}$ and (f) ratio between the integrated intensities of the 2D-and G-mode features $I_\tr{2D}/I_\tr{G}$ along the line scan.}
\label{FigSI7}
\end{center}
\end{figure}

\begin{figure}[!tbh]
\begin{center}
\includegraphics[width=0.65\linewidth]{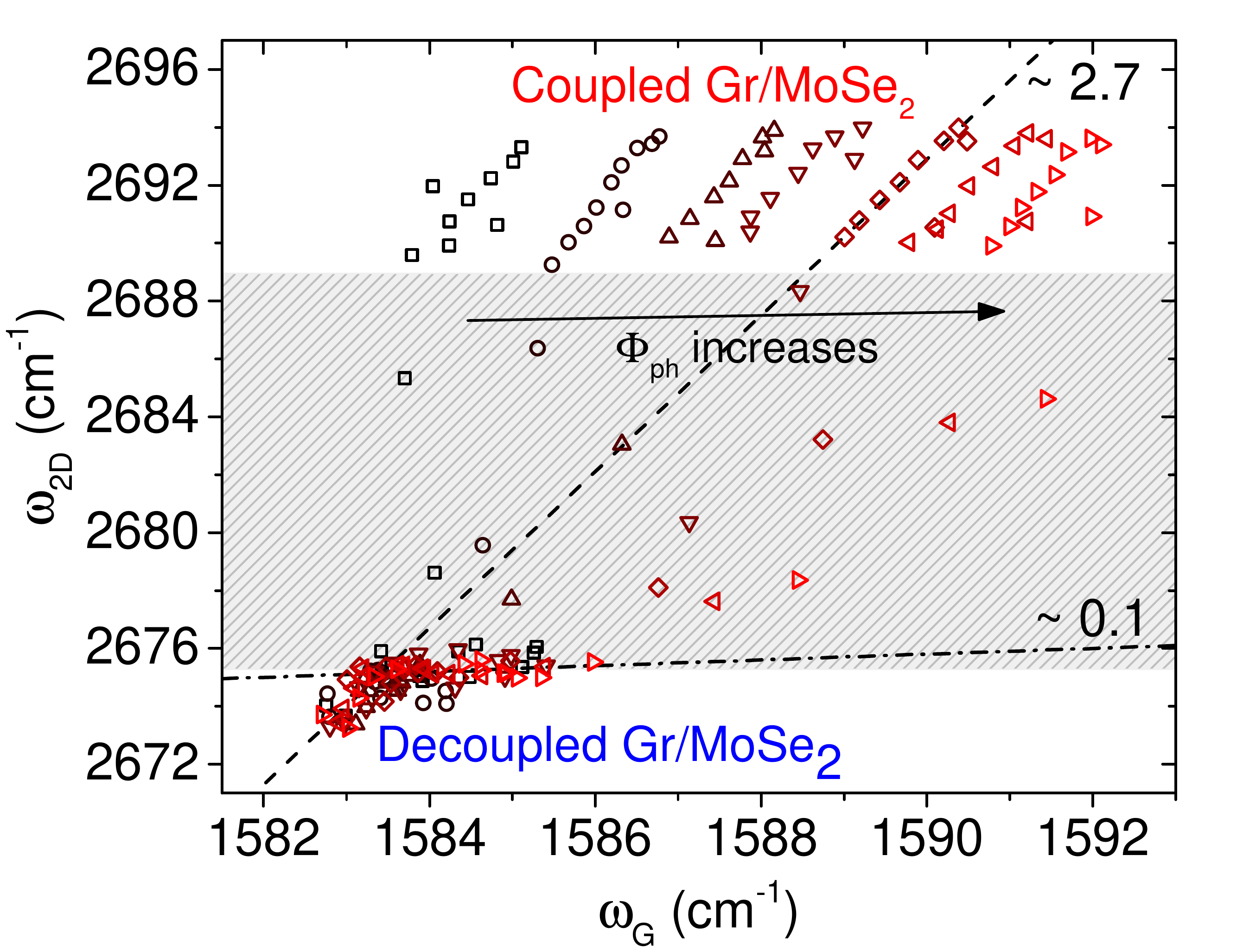}
\caption{Correlation between the 2D- and G-mode frequencies for the same line scans as in Fig.~\ref{FigSI3}. The gray hatched area corresponds to the transition between decoupled and coupled Gr/MoSe$_2$. We notice that for decoupled Gr/MoSe$_2$, the $(\omega_{\rm 2D},\omega_{\rm G})$ points partly collapse onto a same line of slope $\approx 2.7$ for all $\Phi_\tr{ph}$, while another set of points follows a linear correlation with a much reduced slope ($\approx 0.1$), typical from a slight electron doping~\cite{Froehlicher2015a}. In the coupled region, the $(\omega_{\rm 2D},\omega_{\rm G})$ points follow lines, again with a slope $\approx 2.7$ that horizontally shift to higher $\omega_\tr{G}$ for increasing $\Phi_\tr{ph}$. These observations are consistent with the conclusions drawn from the analysis of the Raman maps (see Figs. 1 and \ref{FigSI3}) and from the data in Fig. 2. This horizontal shift corresponds to the increase of doping with $\Phi_\tr{ph}$. The slope of $\approx 2.7$ is in qualitative agreement with the typical slope of $\approx 2.2$ measured for graphene under biaxial strain~\cite{Metten2014}, and suggests negligible contributions from inhomogeneous doping in this restricted region (note that fingerprints of inhomogneous doping are observed on the extended maps shown in Fig.~\ref{FigSI3}).}
\label{FigSI8}
\end{center}
\end{figure}

\clearpage

\section{Additional results obtained on other samples}

\subsection{Interlayer charge transfer in a Gr/MoSe$_2$ heterostructure with an initially hole-doped graphene layer}

\begin{figure}[!tbh]
\begin{center}
\includegraphics[width=\linewidth]{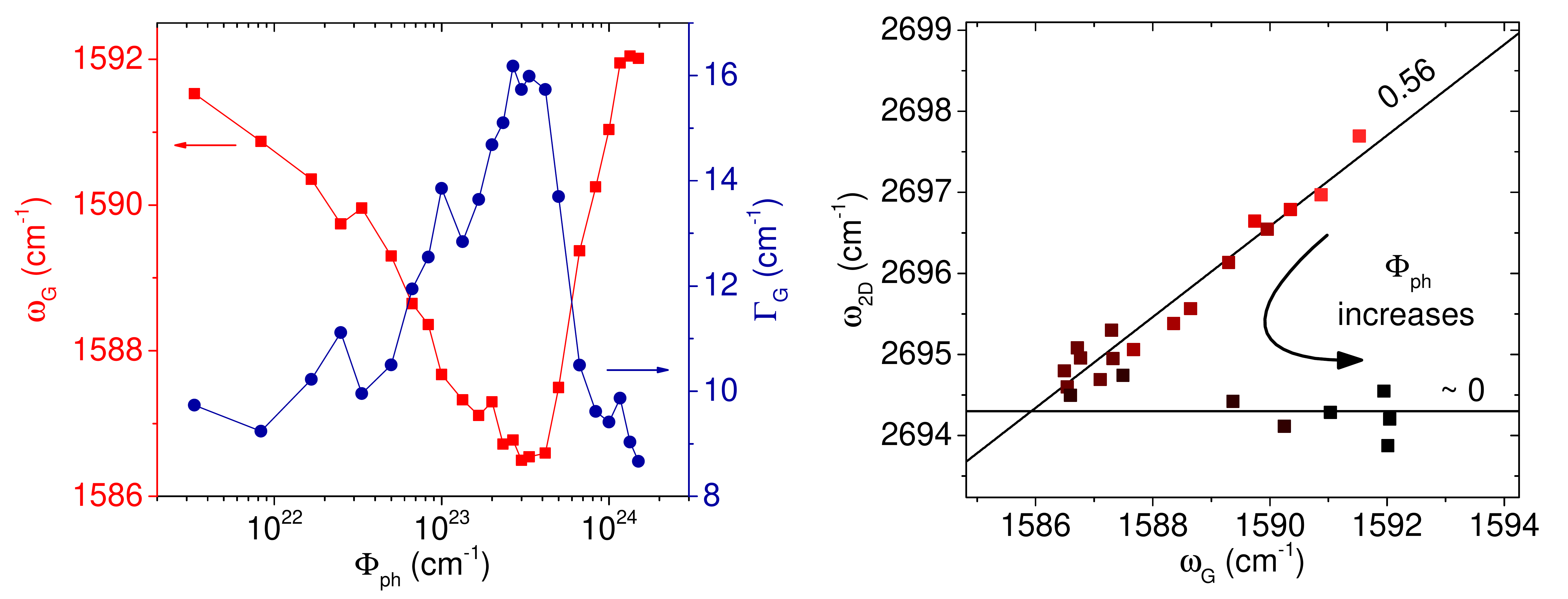}
\caption{(a) Frequency $\omega_\tr{G}$ (red circles and lines, left axis) and FWHM $\Gamma_\tr{G}$ (blue circles and lines, right axis) of the G-mode feature measured on Sample~S$_3$ (see Fig.~4a) at 2.33~eV under ambient conditions as a function of the incident photon flux $\Phi_\tr{ph}$. Lines are guides to the eye. (b) Correlations between the frequencies of 2D- and G-mode features under increasing photon flux $\Phi_\tr{ph}$. We observe a clear linear correlation along two lines with different slopes. At low $\Phi_{\rm ph}$, the frequencies follow a line of slope $0.56$ corresponding to hole doping~\cite{Froehlicher2015a}, while after crossing the charge neutrality point, the frequencies are aligned along a quasi-horizontal line corresponding to electron doping~\cite{Froehlicher2015a} (see also Fig.~2f). As a result, the graphene flake is initially hole-doped and photoexcited electrons are transferred from MoSe$_2$ to graphene.}
\label{FigSI9}
\end{center}
\end{figure}
\clearpage

\subsection{Laser-assisted desorption of molecular adsorbates under high vacuum}

\begin{figure}[!tbh]
\begin{center}
\includegraphics[width=0.55\linewidth]{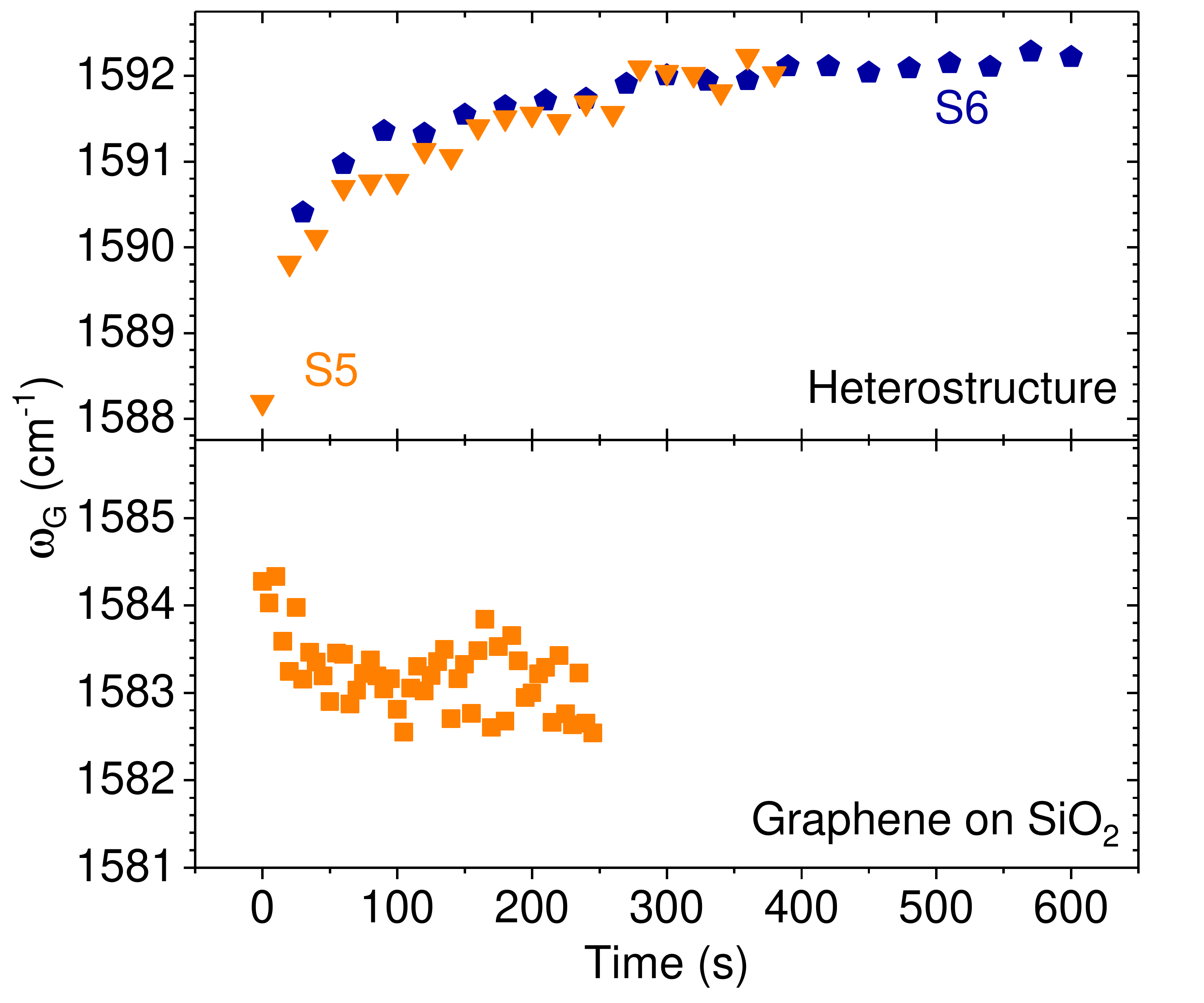}
\caption{Frequency of the Raman G mode ($\omega_{\rm G}$) measured as a function of time under high vacuum on MoSe$_2$/Gr/SiO$_2$ (sample S$_5$), Gr/MoSe$_2$/SiO$_2$ (sample S$_6$) (upper panel) and on a reference graphene monolayer sample on SiO$_2$ (lower panel). Mesusurements were performed at $\Phi_{\rm ph}\sim 5\times 10^{23} ~\rm cm^{-2}~s^{-1}$. The samples had not been illuminated before the measurements.}
\label{FigSI11}
\end{center}
\end{figure}

\clearpage

\subsection{Photoluminescence quenching on various samples}

\begin{figure}[!tbh]
\begin{center}
\includegraphics[width=\linewidth]{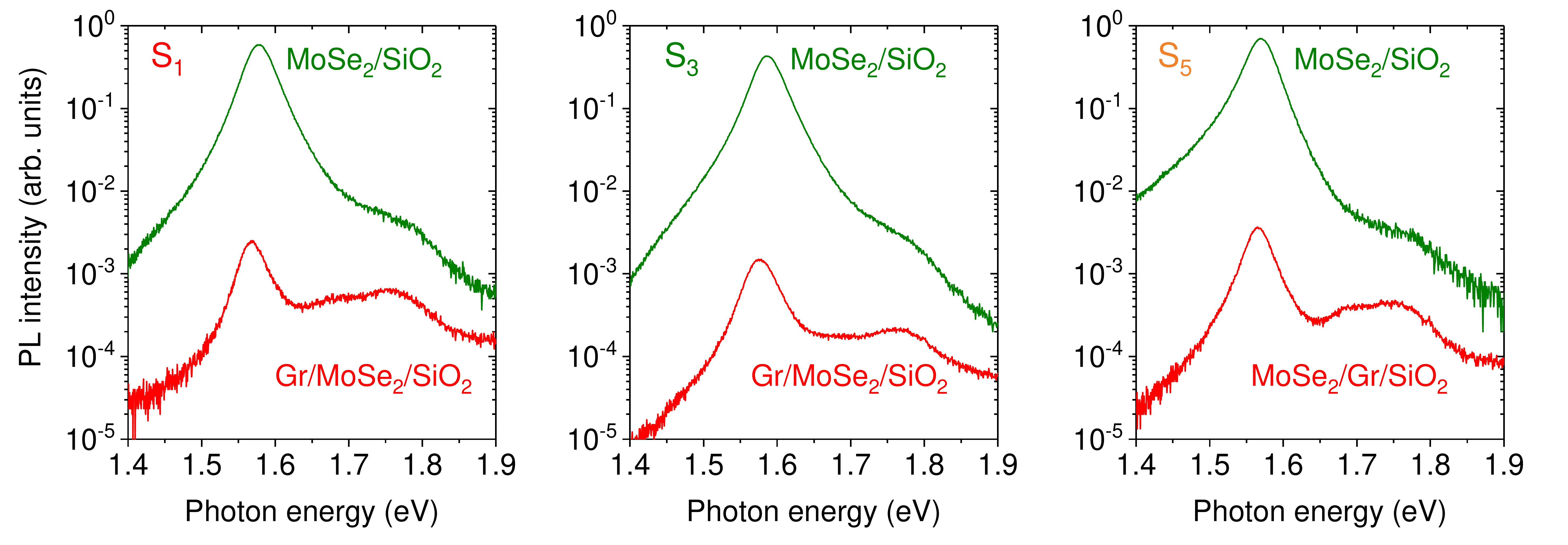}
\caption{Comparison between the photoluminescence spectra recorded on MoSe$_2$/SiO$_2$ and coupled Gr/MoSe$_2$ in ambient air at low $\Phi_{\rm ph}<10^{21} ~\rm cm^{-2}~s^{-1}$ on samples S$_1$, S$_3$, and S$_5$. Very similar quenching factors and PL lineshapes are observed. Note that sample S$_5$ is an \textit{inverted} MoSe$_2$/Gr/SiO$_2$ heterostructure.}
\label{FigSI12}
\end{center}
\end{figure}

\subsection{Exciton dynamics in ambient air and in vacuum}

\begin{figure}[!tbh]
\begin{center}
\includegraphics[width=0.5\linewidth]{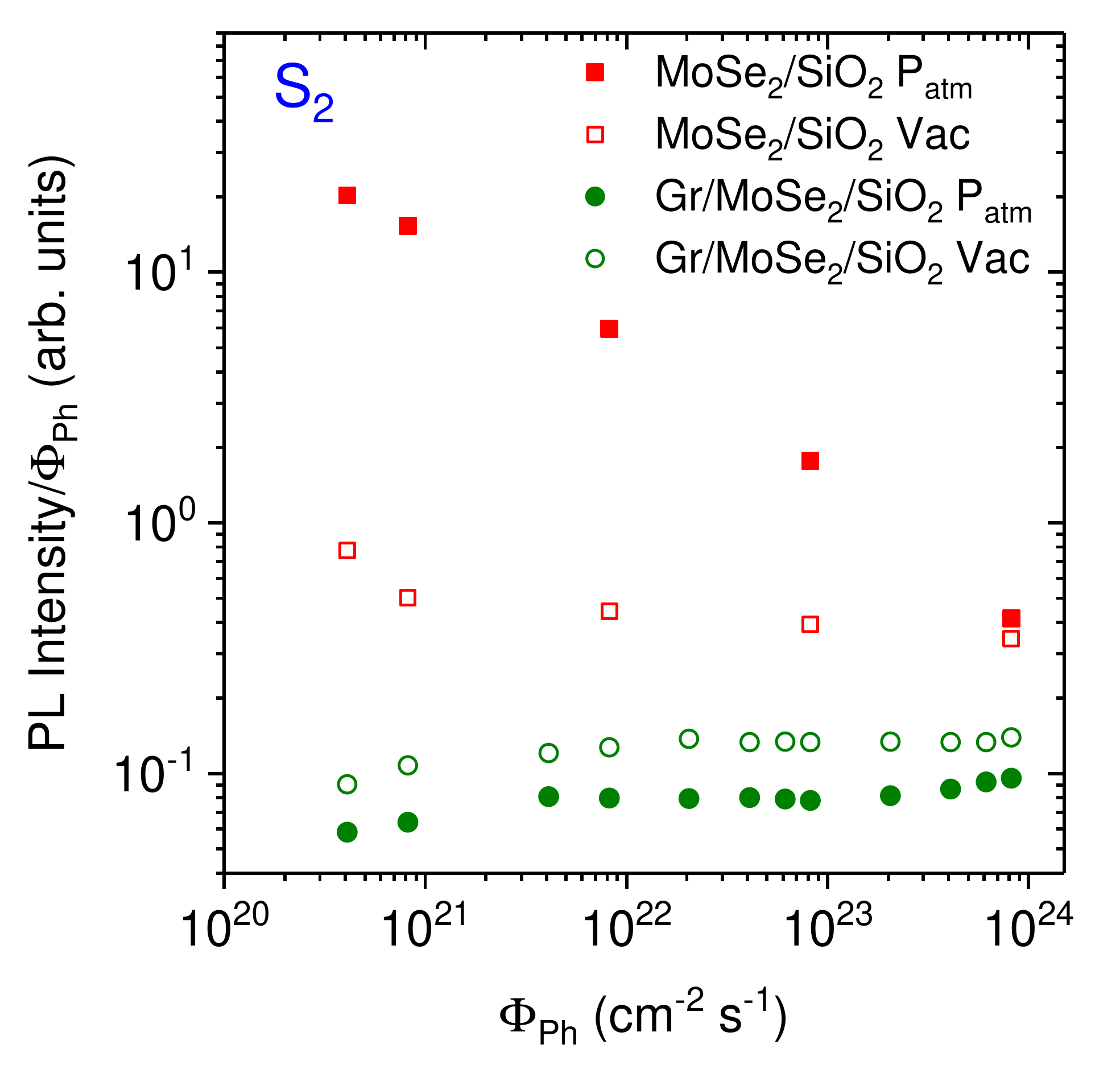}
\caption{Integrated photoluminescence intensity normalized by $\Phi_{\rm ph}$  recorded on Gr/MoSe$_2$/SiO$_2$ (sample S$_2$) in ambient air (filled symbols) and in high vacuum (open symbols) as a function of $\Phi_{\rm ph}$.}
\label{FigSI13}
\end{center}
\end{figure}

\clearpage
\subsection{Comparison between photoinduced doping and exciton dynamics}

\begin{figure}[!tbh]
\begin{center}
\includegraphics[width=\linewidth]{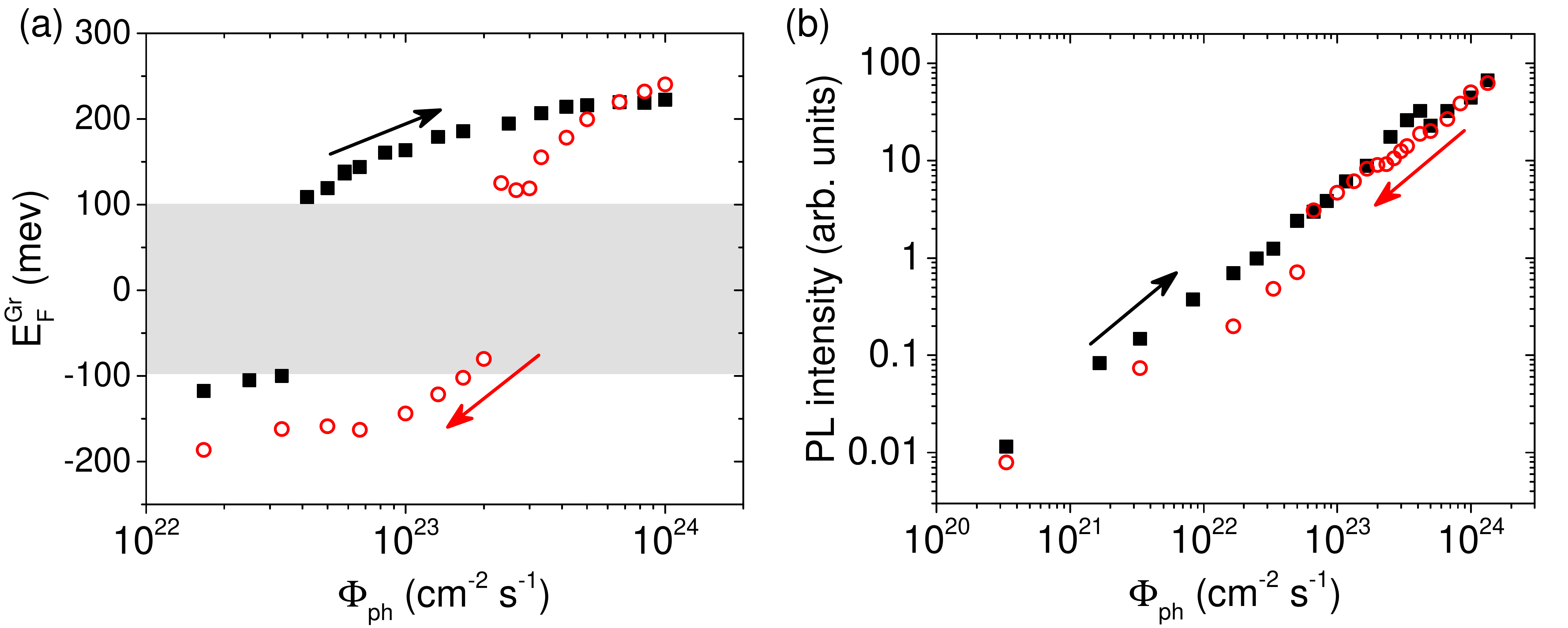}
\caption{(a) Fermi energy in graphene and (b) integrated intensity of the MoSe$_2$ A exciton photoluminescence as a function of the $\Phi_{\rm ph}$ for a forward (black squares) and backward (open red circles) sweep of $\Phi_{\rm ph}$. The measurements were recorded on sample S$_3$.}
\label{FigSI14}
\end{center}
\end{figure}
 
Figure~\ref{FigSI14} shows PL and Raman measurements recorded in ambient air on Sample $S_3$ (see also Fig.~\ref{FigSI9}) along a forward sweep followed by a backward sweep of $\Phi_{\rm ph}$. As opposed to most samples studied in this work, the graphene layer is $p$-doped at low $\Phi_{\rm ph}$ and we clearly see that $E_{\rm F}^{\rm Gr}$ (extracted following the procedure described in the text) has a hysteretic behavior that we attribute to laser-assisted adsorption of electron trapping molecules, such as water or molecular oxygen~\cite{Mitoma2013}. Remarkably, the (linear) evolution of the PL intensity is non-hysterietic, and thus largely independent on the equilibrium value of $E_{\rm F}^{\rm Gr}$ obtained at a given $\Phi_{\rm ph}$. These results further confirm that the ICT processes are likely not solely responsible for the massive PL quenching in Gr/MoSe$_2$, and that molecular adsorbates to affect the charge transfer dynamics.


\clearpage

\subsection{Measurements under quasi-resonant optical excitation}


\begin{figure}[!tbh]
\begin{center}
\includegraphics[width=0.9\linewidth]{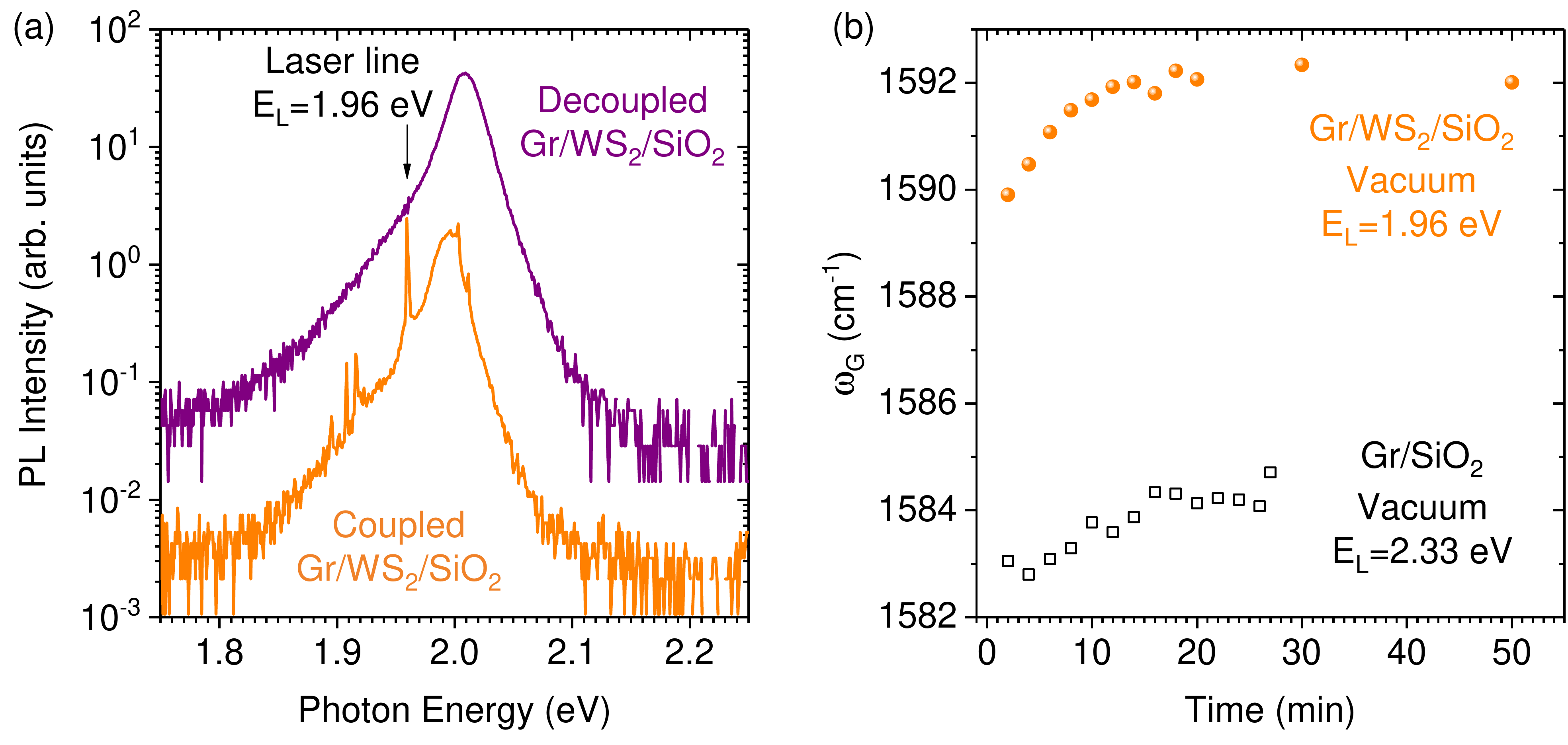}
\caption{PL and Raman measurements on a Gr/WS$_2$/SiO$_2$ heterostructure. The PL spectra in (a) are recorded at $\Phi_\tr{ph}<10^{20}~\tr{cm}^{-2}~\tr{s}^{-1}$ using a laser photon energy $\rm E_L=1.96~\tr{eV}$ slightly below the optical bandgap of WS$_2$~\cite{Li2015}. In these conditions, A excitons only can be formed by means of an upconversion process~\cite{Jones2016,Chervy2017}. Strong PL quenching is observed when comparing the PL intensity from coupled Gr/WS$_2$/SiO$_2$ (orange) to the PL intensity PL from a nearby decoupled Gr/WS$_2$/SiO$_2$ region (purple). The sharp lines above (below) the laser line in the coupled Gr/WS$_2$/SiO$_2$ spectra correspond to the anti-Stokes (Stokes) Raman modes of WS$_2$. (b) Evolution of the Raman G-mode frequency $\omega_{\tr G}$ measured in vacuum as a function of time (similar to Fig.~\ref{FigSI11}) on coupled Gr/WS$_2$/SiO$_2$ at $\rm E_L=1.96~\tr{eV}$ ($\Phi_\tr{ph}\sim 5\times 10^{23}~\tr{cm}^{-2}~\tr{s}^{-1}$). For comparison, a reference measurement performed on the same sample on a Gr/SiO$_2$ region is also shown. The upshifted $\omega_\tr{G}$ relative to this reference is a clear fingerprint of photoinduced doping, as it has been thoroughly discussed on Gr/MoSe$_2$.}
\label{FigSI15}
\end{center}
\end{figure}
\clearpage

\section{Discussion on the frequency of the 2D-mode feature}

\begin{figure}[!tbh]
\begin{center}
\includegraphics[width=0.5\linewidth]{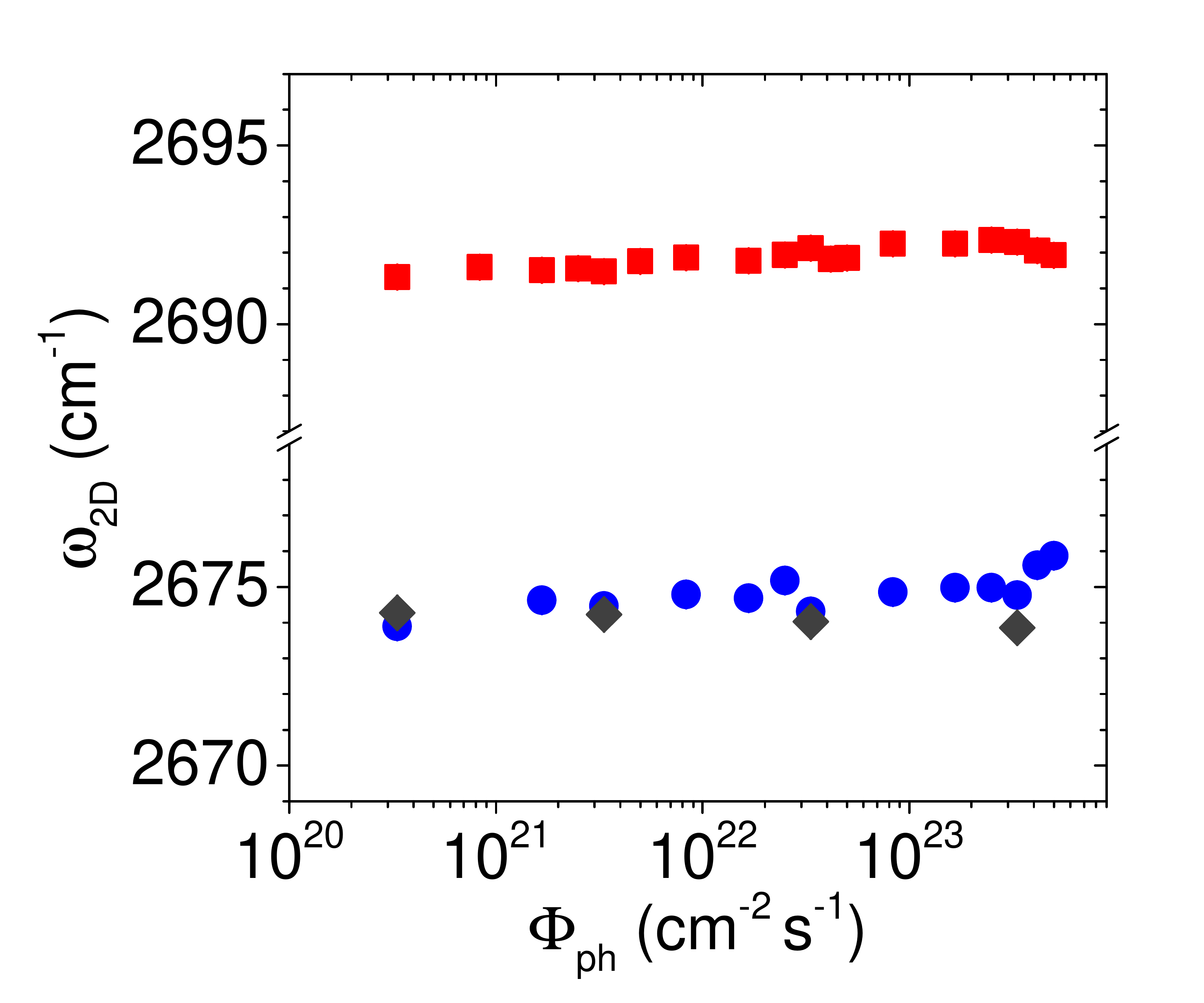}
\caption{2D-mode frequency $\omega_\tr{2D}$ measured in ambient air as a function of $\Phi_\tr{ph}$ on sample $S_1$.}
\label{FigSI10}
\end{center}
\end{figure}

In this section, we briefly comment on the rigid upshift of the 2D-mode frequency observed in coupled Gr-MoSe$_2$. Figs.~1(i) and~2(d) in the main text, and Figs.~\ref{FigSI2}-\ref{FigSI5} reveal a rigid upshift of $\approx 15~\tr{cm}^{-1}$ in coupled Gr/MoSe$_2$ as compared to Gr/SiO$_2$ and decoupled Gr/MoSe$_2$. This upshift cannot be explained by a change of doping~\cite{Froehlicher2015a}. The 2D mode shows more sensitivity to mechanical strain than to doping. However, an upshift of the 2D-mode frequency of around $15~\tr{cm}^{-1}$ caused by strain would also lead to a G-mode upshift of around $7~\tr{cm}^{-1}$~\cite{Lee2012,Metten2014}, irrespective of $\Phi_{\rm ph}$. Such a shift is clearly not observed in all the figures cited above. Interestingly, a similar upshift of the 2D-mode feature has been observed in graphene deposited of thick boron nitride (BN) flakes~\cite{Ahn2013,Forster2013}. For Gr/BN, the 2D-mode upshift has been tentatively explained by dielectric screening due to the thick BN substrate, which reduces the electron-phonon coupling at the $K$ and $K'$ points. It is not obvious that a similar explanation could hold for Gr/single-layer TMD because of the atomic thickness of the TMD. Since the 2D-mode feature interweaves the electron and phonon dispersions~\cite{Maultzsch2004,Basko2008,Venezuela2011,Ferrari2013,Berciaud2013}, another possible explanation could be the modification of the graphene band structure due to van der Waals coupling to MoSe$_2$. However, in the case of MoS$_2$/SLG, it has been calculated that the effects of the interaction on graphene band structure at $\Gamma$, $K$ and $K'$ can be neglected~\cite{Komsa2013,Pierucci2016b}. This intriguing observation of significant 2D-mode stiffening in vdWH will need further theoretical investigations to be fully understood.


\section{Discussion on optical interference effects}
Optical interferences are known to affect the PL and Raman scattering response of 2DM deposited on layered substrates such as Si/SiO$_2$~\cite{Yoon2009,Li2012b,Buscema2014,Froehlicher2016}. Here, we calculated a PL enhancement of only $5~\%$ for air/MoSe$_2$/SiO$_2$/Si as compared to air/Gr/MoSe$_2$/SiO$_2$/Si. This value is much too low to explain the observed PL quenching. We also calculated that for Gr/MoSe$_2$, optical interference effects lead to a negligible enhancement of $I_\tr{2D}/I_\tr{G}$ by about $4~\%$ as compared to the case of Gr/SiO$_2$.


\end{document}